\documentclass[a4paper,12pt]{article}
\usepackage{graphicx}
\usepackage{epsfig}
\usepackage{amsmath}
\usepackage{amssymb}
\usepackage{mathtools}
\DeclarePairedDelimiter{\ceil}{\lceil}{\rceil}

\newcommand{\ds}{\displaystyle }

\title{Continuous time random walk and diffusion with generalized fractional Poisson process}
\author{ {\sl Thomas M. Michelitsch$^{1}$\footnote{Corresponding author,
e-mail~: michel@lmm.jussieu.fr },
Alejandro P. Riascos$^2$ }
\\ \\
$^1$ Sorbonne Universit\'e \\ Institut Jean le Rond d'Alembert, CNRS UMR
7190 \\
4 place Jussieu, 75252 Paris cedex 05, France
\\ \\
$^2$ Instituto de F\'isica, Universidad Nacional Aut\'onoma de M\'exico, \\
Apartado Postal 20-364, 01000 Ciudad de M\'exico, M\'exico
\\ \\ \\ \\
{\it Accepted for publication in Physica A}
}
\begin{document}

\maketitle

\begin{abstract}
A non-Markovian counting process, the
`generalized fractional Poisson process' (GFPP) introduced by Cahoy and Polito in 2013 is analyzed. The GFPP contains two index parameters $0<\beta\leq 1$, $\alpha >0$ and a time scale parameter. Generalizations to Laskin's fractional Poisson distribution and to the fractional Kolmogorov-Feller equation are derived. 
We develop a continuous time random walk 
subordinated to a GFPP in the infinite integer lattice $\mathbb{Z}^d$.
For this stochastic motion, we deduce a `generalized fractional diffusion equation'.
In a well-scaled diffusion limit this motion is governed by the same 
type of fractional diffusion equation as with the fractional Poisson process exhibiting 
subdiffusive $t^{\beta}$-power law for the mean-square displacement.
In the special cases $\alpha=1$ with $0<\beta<1$ the equations of the Laskin fractional Poisson process
and for $\alpha=1$ with $\beta=1$ the classical equations of the standard 
Poisson process are recovered.
The remarkably rich dynamics introduced by the GFPP opens a wide field of applications 
in anomalous transport and in the dynamics of complex systems.
\end{abstract}

{\it Keywords \\  Renewal process, fractional Poisson process and distribution, 
fractional Kolmogorov-Feller equation,\\ continuous time random
walk, generalized fractional diffusion.}

\section{Introduction}

In many complex systems, one has empirically observed asymptotic power laws
instead of the classical exponential patterns. One finds such characteristic behavior for instance in anomalous transport 
and diffusion
\cite{Zaslavsky2002,Shlesinger2017,SaichevZaslavski1997,Gorenflo2007,MetzlerKlafter2000,MetzlerKlafter2004,BarkaiMetzlerKlafter2000}. 
In the description of these phenomena the celebrated Montroll-Weiss 
continuous time random walk (CTRW) model \cite{Shlesinger2017,MontrollWeiss1965,ScherLax1973,KutnerMasoliver1990} has become a major approach where the empirical power laws 
lead to descriptions that involve fractional calculus.
Meanwhile, the CTRW of Montroll and Weiss has been 
applied to a vast number of problems in various contexts especially to model anomalous diffusion 
and transport phenomena
\cite{Shlesinger2017,KutnerMasoliver1990,ZumofenShlesingerKlafter1996} including stochastic processes 
in protein folding \cite{Brungelson1989}, 
the dynamics of chemical reactions \cite{SungBarkaiSilbey2002},
transport phenomena in low dimensional chaotic systems \cite{SolomonSwinner1993}, 
`Aging Continuous Time Random Walks' \cite{MonthusBouchaud1996,BarkaiCheng2003},
CTRW with correlated memory kernels have been introduced \cite{ChechkinHofmannSokolov2009}, 
and many further applications of the CTRW exist. In the classical picture, CTRWs are random walks subordinated to a Poisson renewal process 
with exponentially decaying 
waiting time densities
and the Markovian property \cite{MetzlerKlafter2000,MetzlerKlafter2004,Feller1971}. 
However, one has found that many complex systems 
are rather governed by non-Markovian fat-tailed waiting time densities not compatible with the 
classical Poisson process picture
\cite{SaichevZaslavski1997,MetzlerKlafter2000,Laskin2003,MainardiGorenfloScalas2004}.
Based on a renewal process to our present knowledge first introduced by Repin and Saichev \cite{RepinSaichev2000}, Laskin developed a non-Markovian fractional generalization 
of the Markovian Poisson process. He called this process `{\it fractional Poisson process}' \cite{Laskin2003}
and demonstrated various important applications \cite{Laskin2009}. 
The fractional Poisson process is of utmost importance as it produces power-law patterns as observed in 
anomalous subdiffusion due to fat-tailed waiting time densities. Meanwhile, a large literature exists that
analyze fractional Poisson processes and related random motions
\cite{Gorenflo2007,MetzlerKlafter2004,BeghinOrsinger2009,GorenfloMainardi2006,Gorenflo2010,GorenfloMainardi2013,MeerschaertEtal2011}.
In the present paper, our aim is the development of a stochastic motion based on a generalization
of the Laskin's fractional Poisson process. This process was introduced by Cahoy and Polito in 2013 \cite{PolitoCahoy2013} and several features were analyzed in that paper. 
In the present paper, our goal is to complement that analysis and to investigate related stochastic motions.
We call this renewal process `{\it generalized fractional Poisson process}' (GFPP). 
Both the GFPP like the fractional Poisson process are 
non-Markovian and introduce long-memory effects.
We will come back to this issue subsequently.
The present paper is organized as follows. In the first part (Section \ref{CTRW}) 
we briefly outline general concepts of renewal theory and the CTRW approach by Montroll and Weiss 
\cite{MontrollWeiss1965,ScherLax1973}.
In Section \ref{FPP} we briefly discuss essential features 
of Laskin's fractional Poisson counting process and of the fractional Poisson distribution. In Section \ref{generalized} we generalize the Laskin process and 
define the `{\it generalized fractional Poisson process}' (GFPP). We derive 
its waiting time density and related distributions.
In Section \ref{GeneralizedFractPoissonDistri} the probability 
distribution of $n$ arrivals in a GFPP is deduced. We refer this 
distribution to as `{\it generalized fractional Poisson distribution}' (GFPD).
We derive in Section \ref{GenFractionalPoissonProcess} 
the expected number of arrivals in a GFPP and analyze its asymptotic power law features. 
The GFPP contains two index parameters $0<\beta\leq 1$ and $\alpha>0$ and 
one additional parameter ($\xi^{-\frac{1}{\beta}}$) that defines a characteristic time scale of the process. 
The Laskin fractional Poisson process 
is a special case for $\alpha=1$ and $0<\beta<1$ of the GFPP,
as well as the standard Poisson process (recovered for $\alpha=1$ and $\beta=1$).
Further the case $\beta=1$ and $\alpha>0$ produces an Erlang type process.
As an application, in Sections \ref{CTRWnetworks}-\ref{genfractdiffusion}, we consider
a random walk subordinated to a GFPP on undirected 
networks and multidimensional infinite lattices in the framework of 
Montroll-Weiss CTRW. We derive `generalized fractional Kolmogorov-Feller equations' and for the resulting
diffusion process a `generalized fractional diffusion equation'.
We show explicitly that for $\alpha=1$ ($0<\beta<1$) these equations reproduce their 
fractional counterparts \cite{SaichevZaslavski1997}
of the Laskin fractional Poisson process and for $\alpha=1$, $\beta=1$ 
the classical equations of standard Poisson. Finally we analyze the `well-scaled' diffusion limit and obtain the same type of fractional diffusion equation 
as in walks subordinated to Laskin's fractional Poisson process.
As a major analytical tool in this paper, we utilize Laplace transforms
where we deal with {\it causal} non-negative probability distributions (non-negative probability measures) in 
the distributional 
sense of Gelfand and Shilov generalized functions \cite{GelfangShilov1968}. The Laplace 
transform ${\tilde f}(s)$ of a {\it causal function} $f(t)=\Theta(t){\bar f}(t)$ 
is defined by, e.g. \cite{GelfangShilov1968}\footnote{We denote as
$\int_{0_{-}}(\ldots) = \lim_{\epsilon\rightarrow 0+} \int_{-\epsilon}(\ldots)$.}
\begin{equation}
 \label{LaplaceDef}
 {\cal L}\{f(t)\}=:{\tilde f}(s)= 
 \int_{0_{-}}^{\infty}  e^{-st} \Theta(t){\bar f}(t){\rm d}t 
 ,\hspace{1cm} s=\sigma+i\omega ,\hspace{1cm} \sigma >\sigma_0,
\end{equation}
where $\Theta(t)$ indicates the Heaviside step 
function defined by $\Theta(t)= 1$ for $t\geq 0$ and $\Theta(t)=0$ for $t<0$, especially $\Theta(0)=1$.
However, we emphasize that $\Theta(0_{-})=\lim_{\epsilon\rightarrow 0+}\Theta(-\epsilon)=0$ ($\epsilon >0$) 
is vanishing as the lower integration limit since $0_{-} = -\epsilon$ ($\epsilon\rightarrow 0+$) 
is infinitesimally negative.
The essential point is that the Laplace transform captures the entire nonzero contributions 
of a causal distribution $f(t)$. We utilize throughout 
this paper definition
(\ref{LaplaceDef}) for Laplace transforms since it especially 
convenient when dealing with causal probability measures and causal operators.
The Laplace inversion is performed by
\begin{equation}
 \label{inverseLpalace}
 \begin{array}{l}
 \ds f(t)= {\bar f}(t)\Theta(t)= {\cal L}^ {-1}\{{\tilde f}(s)\} = \frac{1}{2\pi i} 
\ds \int_{\sigma -i\infty}^{\sigma+i\infty}e^{st} {\tilde f}(s){\rm d}s ,\hspace{1cm} \Re(s)=\sigma >\sigma_0 ,\\ \\
\ds  \hspace{3cm}= \frac{e^{\sigma t}}{2\pi}\int_{-\infty}^{\infty} 
e^ {i\omega t}{\tilde f}(\sigma+i\omega){\rm d}\omega
\end{array}
 \end{equation}
with suitably chosen $\Re{(s)}=\sigma$. We only 
mention here the important property for causal functions
\begin{equation}
 \label{it-follows-that}
 \begin{array}{l}
 \ds {\cal L}\left\{\frac{d^m}{dt^m}(\Theta(t){\bar f}(t))\right\}  = s^m {\tilde f}(s) - \underbrace{\sum_{k=0}^{m-1}s^{m-1-k}\frac{d^k}{dt^k}\left(f(t)\Theta(t)\right)\Big|_{t=0_{-}}}_{=0} \\ \\
 \ds {\cal L}^{-1}\{ s^m {\tilde f}(s)\} = \frac{d^m}{dt^m}\left\{\Theta(t){\bar f}(t)\right\} ,\hspace{1cm} m \in \mathbb{N}
 \end{array}
\end{equation}
i.e. Laplace inversion of $s^m {\tilde f}(s)$ captures 
in the derivatives the 
Heaviside step function\footnote{We notice the distributional representation (See Appendix \ref{AppendLaplacetrafo}):\\
${\cal L}^{-1}\left\{ s^m {\tilde f}(s) \right\}= \frac{d^m}{dt^m}\left({\bar f}(t)\Theta(t)\right) = 
\Theta(t)\frac{d^m}{dt^m}{\bar f}(t) + \sum_{k=1}^m\frac{m!}{k!(m-k)!}\delta^{(k-1)}(t)
\frac{d^{m-k}}{dt^{m-k}}{\bar f}(t)$, where is denoted $\delta^{(s)}(t) =\frac{d^s}{dt^s}\delta(t)$.}. 
This allows covering causal distributions that exhibit discontinuities especially when they are `switched on' 
at $t=0$.
We emphasize property (\ref{it-follows-that}) as it allows us later to determine the appropriate 
fractional derivative kernels (See especially Appendix \ref{AppendLaplacetrafo}).
In our demonstration when no derivatives are involved we often skip the Heaviside $\Theta(t)$ step
function in expressions of causal
distributions. However, for clarity, we
include the notation ``$\Theta(t)$'' in all cases such as (\ref{it-follows-that}) where time derivatives are involved.
We mention further with the above definition of Laplace transform (\ref{LaplaceDef}), 
that the Dirac's $\delta$-function is entirely captured. This is expressed by ${\cal L}\{\delta(t)\}=1$.
Further properties are discussed in Appendix \ref{AppendLaplacetrafo}.
We avoided in the present paper too extensive
mathematical derivations which can be found in a complementary paper 
\cite{MichelitschRiascosGFPP2019}. 

Contrary to the version of the present paper to appear in {\it Physica A}, in the present version we have analyzed in Section \ref{genfractdiffusion}
a `well-scaled' diffusion limit which yields the time fractional diffusion equation (87.B).

\section{Renewal process and CTRW} 
\label{CTRW}

Let us briefly outline some basic features of 
renewal processes and waiting time distributions (See e.g. \cite{Gorenflo2007}) 
as a basis of the CTRW approach of Montroll and Weiss \cite{MontrollWeiss1965}.
We assume that a walker makes jumps at random times $0\leq t_1,t_2,\ldots t_n,\ldots \infty$. 
The non-negative random times when jumps occur we also refer to as `arrival times' $t_k$.
We start the observation of the walk at time $t=0$ where the walker is sitting on his initial 
position until making the first jump at $t=t_1$ (arrival of the first jump event) and so forth.
The waiting times $\Delta t_k = t_k-t_{k-1} \geq 0$ between two successive (jump-) 
events are assumed to be drawn for each jump from the same 
probability {\it density} distribution (PDF) $\chi(t)$ which we refer 
to as `{\it waiting time density}' or `{\it jump density}'.
The waiting times $\Delta t_k$ then
are independent and 
identically distributed (IID) random variables.
$\chi(\tau) {\rm d}\tau $ indicates the 
probability that the first jump event (first arrival) happens exactly at time $\tau$.
The probability that the walker at least has jumped once within the time interval $[0,t]$ is given by
the cumulative probability distribution
\begin{equation}
 \label{waitintimedis}
prob(\Delta t \leq t)= \Psi(t) = \int_0^t \chi(\tau){\rm d}\tau ,\hspace{0.5cm} t\geq 0 , 
\hspace{0.5cm} \lim_{t\rightarrow\infty} \Psi(t) =1-0
\end{equation}
with initial condition $\Psi(t=0)=0$. The probability that the walker within $[0,t]$ still is 
waiting on its departure site is given by
\begin{equation}
 \label{waiting-timeprob}
 prob(\Delta t \geq t) = \Phi^{(0}(t)=1-\Psi(t) =\int_t^{\infty}\chi(\tau){\rm d}\tau .
\end{equation}
The cumulative probability distribution (\ref{waiting-timeprob}) that the walker during $[0,t]$ still is waiting on
its initial site also is called `{\it survival probability}' and indicates the probability
that the waiting time $\Delta t \geq t$. We notice that the jump densities\footnote{We abbreviate 
probability {\it density} functions or short `densities' by `PDF'.} such as $\chi(t)$ have dimension $[\mathrm{time}]^{-1}$
whereas the cumulative distributions  (\ref{waitintimedis}), (\ref{waiting-timeprob}) 
are dimensionless probabilities. In addition, the waiting time density $\chi(\tau)$ is
normalized (see (\ref{waitintimedis})) which is reflected by the property ${\tilde \chi}(s)|_{s=0}=1$ 
of its Laplace transform.
It follows that the survival probability (\ref{waiting-timeprob})
tends to zero as time approaches infinity $\lim_{t\rightarrow\infty} \Phi^{(0}(t) \rightarrow 0$.
The PDF $\chi^{(n)}(t)$ that the walker makes its
$n$th jump {\it exactly at time $t$} 
then is\footnote{We refer this density also to as {\it $n$-jump density}.}
\begin{equation}
 \label{probofnjumps}
 \chi^{(n)}(t) = \int_0^{\infty}\ldots \int_0^{\infty}\chi(\tau_1)\ldots 
 \chi(\tau_n)\delta\left(t-\sum_{j=1}^n\tau_j)\right){\rm d}\tau_1\ldots {\rm d}\tau_n
\end{equation}
having the Laplace transform
\begin{equation}
\label{laplacenjumps}
\begin{array}{l} 
\displaystyle 
{\tilde \chi}^{(n)}(s) = \int_0^{\infty}{\rm d}\tau_1 \ldots \int_0^{\infty}{\rm d}\tau_n  \chi(\tau_1)\ldots \chi(\tau_n)\int_0^{\infty}{\rm d} t e^{-st} \delta\left(t-\sum_{j=1}^n\tau_j)\right) ,\\ \\
\displaystyle \hspace{0.5cm} = \int_0^{\infty}\ldots \int_0^{\infty}\chi(\tau_1)\ldots \chi(\tau_n) {\rm d}\tau_1\ldots {\rm d}\tau_n e^{-s\sum_{j=1}^n\tau_j} = \left(\int_0^{\infty}\chi(\tau)e^{-s\tau}{\rm d}\tau\right)^n = ({\tilde \chi}(s))^n
\end{array}
\end{equation}
which by putting $s=0$ yields the normalization condition  
$\int_0^ {\infty}{\tilde \chi}^{(n)}(t){\rm d}t =1$ 
as a consequence of the normalization of the one-jump density ${\tilde \chi}(s)|_{s=0} =1$.

The probability that a walker within the time interval $[0,t]$ has made $n=0,1,2,\ldots $ 
steps then is related with (\ref{probofnjumps}) 
by the convolution
\begin{equation}
 \label{nstepprobabilityuptot}
 \Phi^{(n)}(t) = \int_0^t\left(1-\Psi(t-\tau)\right)\chi^{(n)}(\tau){\rm d}\tau = \int_0^t \Phi^{(0)}(t-\tau)\chi^{(n)}(\tau){\rm d}\tau
\end{equation}
where we account for that the walker makes its $n$th jump at a time $\tau<t$ with probability 
$\chi^{(n)}(\tau){\rm d}\tau$ and waits during $t-\tau$ with probability 
$\Phi^ {(0)}(t)=1-\Psi(t-\tau)$ where over all combinations from $\tau=0$ to $t$ is integrated.
The probabilities (\ref{nstepprobabilityuptot}) are dimensionless distributions whereas 
the $n$-jump densities (\ref{laplacenjumps})
have dimension $[\mathrm{time}]^{-1}$.
We mention the normalization condition shown later
\begin{equation}
 \label{phinnormaliszation}
 \sum_{n=0}^{\infty} \Phi^{(n)}(t) = 1 .
\end{equation}
We observe then the recursive relations
\begin{equation}
 \label{nstepssteps}
 \begin{array}{l}
\ds  \chi^{(n)}(t) = \int_0^t \chi^{(n-1)}(\tau)\chi(t-\tau){\rm d}\tau ,\hspace{1cm} \chi^{(0)}(t)=\delta(t) \\ \\
 \ds \Phi^{(n)}(t) = 
 \int_0^t \Phi^{(n-1)}(\tau)\chi(t-\tau){\rm d}\tau  ,\hspace{1cm} \Phi^{(0)}(t)=1-\Psi(t) .
 \end{array}
\end{equation}
We then get for the Laplace transform of (\ref{nstepprobabilityuptot})
\begin{equation}
 \label{laplansteps}
 {\tilde \Phi}^{(n)}(s) =\frac{1-{\tilde \chi}(s)}{s}({\tilde \chi}(s))^n =
 \frac{1}{s}\left({\tilde \chi}^{(n)}(s)-{\tilde \chi}^{(n+1)}(s)\right) , \hspace{1cm} n=0,1,2,\ldots \infty,
\end{equation}
recovering ${\tilde \Phi}_0(s)=\frac{1-{\tilde \chi}(s)}{s}$ when 
we account for ${\chi}_0(t)=\delta(t)$ with ${\tilde \chi}^{(0)}(s) = {\cal L}\{\delta(t)\} = 1$. 
Introducing a generating function in the Laplace domain as\footnote{This series is
converging for $|v|\leq 1$ when $s\neq 0$ and for $|v|<1$ when $s=0$ as consequence of 
$|{\tilde \chi}(s)|\leq 1$ (equality for $s=0$).}
\begin{equation}
 \label{generatingsteps-laplace}
{\tilde  G}(s,v)= \sum_{n=0}^{\infty} {\tilde \Phi}^{(n)}(s)\, v^n = 
\frac{1-{\tilde \chi}(s)}{s} \frac{1}{1-v{\tilde \chi}(s)} ,
 \hspace{1cm} |v|\leq 1
\end{equation}
where ${\tilde G}(s,1)=s^{-1}$ proving then normalization condition (\ref{phinnormaliszation}).
A quantity of great interest in the following analysis is
the (dimensionless) expected number of arrivals ${\bar n}(t)$ (expected number of jumps) 
that happen within the time interval $[0,t]$. This quantity is obtained by
\begin{equation}
 \label{averageNumberSteps}
 \ds {\bar n}(t) = \sum_{n=0}^{\infty} n \Phi^{(n)}(t) = \frac{d}{d v} G(t,v)\Big|_{v=1}
 = {\cal L}^{-1} \left\{  \frac{{\tilde \chi}(s)}{s (1-{\tilde \chi}(s))} \right\} .
\end{equation}
For further discussions of {\it renewal processes} we refer to the references
\cite{Gorenflo2010,GorenfloMainardi2013,Cox1967}.
With these general remarks on renewal processes, let us briefly discuss in the following section 
Laskin's fractional Poisson process in order to generalize this process
in Section \ref{generalized}.

\section{Fractional Poisson distribution}
\label{FPP}
The fractional Poisson process was introduced by Laskin and several aspects of this process  were thoroughly analyzed
in several papers \cite{SaichevZaslavski1997,Laskin2003,MainardiGorenfloScalas2004,Laskin2009}.
The Laskin fractional Poisson process is defined by a renewal process
with a Mittag-Leffler waiting time density\footnote{We employ throughout this paper 
as equivalent notations $\zeta !=\Gamma(\zeta+1)$ for the $\Gamma$-function.}
\begin{equation}
 \label{Mittag-Leffler-waiting-time-PDF}
 \begin{array}{l}
\ds  \chi_{\beta}(t) =
  \sum_{n=1}^{\infty}(-1)^{n-1} \xi^{n} 
\frac{ t^{n\beta-1}}{(n\beta-1)!}  \\ \\
\ds \hspace{0.5cm} = \xi t^{\beta-1} \sum_{n=0}^{\infty}
\frac{(-1)^n \xi^n t^{n\beta}}{\Gamma(\beta n+\beta)}= \xi t^{\beta-1} E_{\beta,\beta}(-\xi t^{\beta}) 
,\hspace{0.5cm} t>0 \\ \\
\hspace{0.5cm} =  \frac{d}{dt}[1-E_{\beta}(-\xi t^{\beta})] 
,\hspace{0.5cm} \xi>0 \hspace{0.5cm} 0<\beta\leq 1 , \hspace{1cm} \xi >0 
\end{array}
\end{equation}
where we introduced the Mittag-Leffler function of order $\beta$ \cite{Gorenflo2007,Laskin2003,Gorenflo2010,GorenfloKilbas2014}
\begin{equation}
 \label{Mittag-L-intro}
E_{\beta}(z) = \sum_{n=0}^{\infty} \frac{z^n}{\Gamma(\beta n+1)}
\end{equation}
and the generalized
Mittag-Leffler function
\begin{equation}
 \label{Mittag-L-intro-gener}
E_{\beta,\gamma}(z) = \sum_{n=0}^{\infty} \frac{z^n}{\Gamma(\beta n+\gamma)}
\end{equation}
where $E_{\beta,1}(z) = E_{\beta}(z)$.
Since the argument $\xi t^{\beta}$ of the Mittag-Leffler function is dimensionless, 
$\xi$ is a characteristic parameter of physical dimension
$[\mathrm{time}]^{-\beta}$. 
It is straight-forward to show that the Mittag-Leffler jump density 
(\ref{Mittag-Leffler-waiting-time-PDF}) 
has the Laplace transform
\begin{equation}
 \label{LaplacetrafoMittag-Letffler}
 \begin{array}{l}
\ds  {\tilde \chi}_{\beta}(s)= \sum_{n=1}^{\infty}(-1)^{n-1} \xi^{n} 
\int_0^{\infty}e^{-st}\frac{ t^{n\beta-1}}{(n\beta-1)!}{\rm d}t = \sum_{n=1}^{\infty}(-1)^{n-1} 
\xi^{n}s^{-n\beta} \\ \\ \ds  \hspace{0.5cm} =  \frac{\xi}{\xi+s^{\beta}} 
, \hspace{1cm} s=\sigma+i\omega ,\hspace{0.5cm} \sigma > \xi^{\frac{1}{\beta}}
\end{array}
\end{equation}
converging for $\sigma >\xi^{\frac{1}{\beta}}$. For $\beta=1$, (\ref{Mittag-Leffler-waiting-time-PDF}) recovers the exponential 
jump density $\chi_1(t)=\xi e^{-\xi t} $ of the standard Poisson process.
The survival probability in the fractional Poisson process
is then of the form of a {\it Mittag-Leffler function}
\begin{equation}
 \label{survival}
\Phi^{(0)}_{\beta}(t)= 1-\Psi_{\beta}(t) = 
{\cal L}^{-1}\left\{\frac{s^{\beta-1}}{\xi+s^{\beta}} \right\}= 
E_{\beta}(-\xi t^{\beta}) ,\hspace{1cm} 0 < \beta\leq 1
\end{equation}
where for $\beta=1$ the standard Poisson process with survival probability $E_1(-\xi t) = e^{-\xi t}$ is recovered. 
The {\it fractional Poisson distribution} is defined as 
the probability $\Phi_{\beta}^{(n)}(t)$ of exactly $n$ arrivals within time interval $[0,t]$ for a renewal process 
with jump density (\ref{Mittag-Leffler-waiting-time-PDF}) and is
obtained as \cite{Laskin2003}

\begin{equation}
 \label{fractionalpoission-distribution}
 \begin{array}{l}
 \ds \Phi^{(n)}_{\beta}(t) = \frac{(\xi t^{\beta})^n}{n!}\frac{{\rm d}^n}{{\rm d}\tau^n}E_{\beta}(\tau)\Big|_{\tau=-\xi t^{\beta}}  \\ \\ \hspace{0.5cm} \ds = 
 \frac{(\xi t^{\beta})^n}{n!} \sum_{m=0}^{\infty}\frac{(m+n)!}{m!}\frac{(-\xi t^{\beta})^m}{\Gamma(\beta(m+n)+1)} ,\hspace{0.25cm} 0<\beta \leq 1 ,
 \hspace{0.25cm} n=0,1,2,\dots 
 \end{array}
\end{equation}
For $\beta=1$ the fractional Poisson distribution and $\alpha=1$, $\beta=1$
recovers the standard Poisson distribution 
\begin{equation}
 \label{standardpoisson}
\Phi^{(n)}_{1}(t)= \Theta(t)\frac{(\xi t)^n}{n!}e^{-\xi t} ,\hspace{1cm} n=0,1,2,\ldots 
\end{equation}
%
%
%

\section{Generalization of the fractional Poisson process}
\label{generalized}

We now analyze a counting process that generalizes the fractional Poisson process. The waiting 
time density of this generalized
process has
the Laplace transform 
\begin{equation}
 \label{jump-gen-fractional-laplacetr}
 {\tilde \chi}_{\beta,\alpha}(s) = \frac{\xi^{\alpha}}{(s^{\beta}+\xi)^{\alpha}} ,\hspace{0.5cm} 
 0<\beta\leq 1 ,\hspace{0.5cm} \alpha >0 , \hspace{1cm} \xi >0
\end{equation}
and was introduced by Cahoy and Polito \cite{PolitoCahoy2013}. The goal of the remaining parts is to complement their analysis of the process and
to develop a CTRW that is governed by this process to derive generalizations of the fractional Kolmogorov- and diffusion equations. 
We see that (\ref{jump-gen-fractional-laplacetr}) 
generalizes (\ref{LaplacetrafoMittag-Letffler}) of the Laskin fractional Poisson process where any
non-negative power $\alpha \in \mathbb{R}_+ $ is admissible in (\ref{jump-gen-fractional-laplacetr}).
We call the renewal process with waiting time density defined by (\ref{jump-gen-fractional-laplacetr})
the {\it generalized fractional Poisson process} (GFPP). The characteristic dimensional constant $\xi$ in
(\ref{jump-gen-fractional-laplacetr}) has physical dimension $[\mathrm{time}]^{-\beta}$ and defines 
a characteristic time scale.
Per construction 
${\tilde \chi}_{\beta,\alpha}(s)|_{s=0}= \int_0^{\infty} \chi_{\beta,\alpha}(t){\rm d}t=1$ 
reflecting normalization of the waiting time density.
The GFPP recovers for $\alpha=1$ , $0<\beta<1$ the fractional Poisson process, for $\alpha>0$, $\beta=1$
an Erlang type process (See \cite{GorenfloMainardi2013}) allowing also fractional powers $\alpha$,
and for $\alpha=1$, $\beta=1$ the standard Poisson process.

We observe in (\ref{jump-gen-fractional-laplacetr}) that the jump density
has for $0<\beta<1$ and for all
$\alpha>0$
diverging mean (diverging expected time for first arrival), namely 
$-\frac{d}{ds}{\tilde \chi}_{\beta,\alpha}(s)|_{s=0}=\int_0^{\infty} t 
\chi_{\beta,\alpha}(t) \rightarrow +\infty$, and a finite mean first arrival time in the Erlang regime
$\alpha \xi^{-1}$ for $\alpha>0$ with $\beta=1$.

\begin{figure*}[!ht]
\begin{center}
\includegraphics*[width=1.0\textwidth]{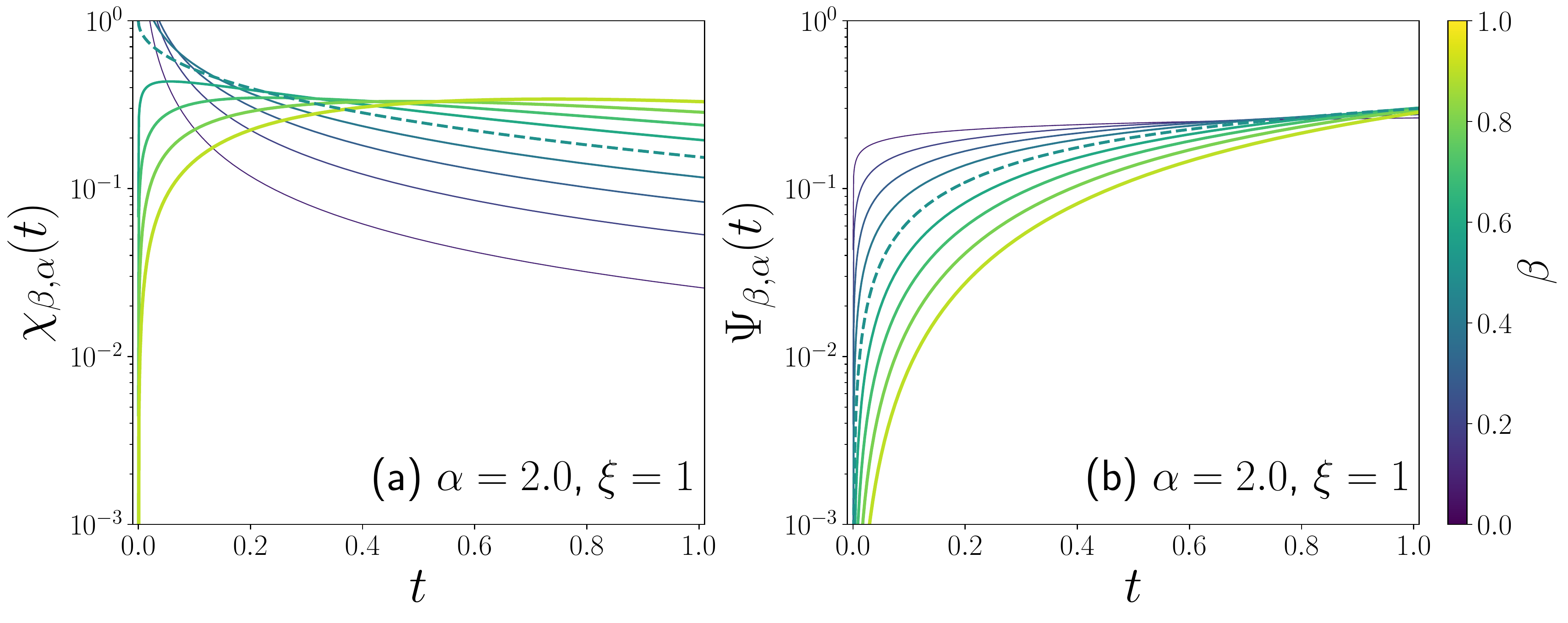}
\end{center}
\vspace{-5mm}
\caption{\label{Fig_1} (a) Jump density $\chi_{\beta,\alpha}(t)$ and (b)
cumulated probability $\Psi_{\beta,\alpha}(t)$  as a function of $t$. 
We explore the case $\alpha=2.0$  for different values  $0< \beta \leq 1$ codified in the colorbar. 
The results were obtained numerically using $\xi=1$ and Eqs. (\ref{Laplainv}) and (\ref{cumulgenfract}). 
We depict with dashed lines the results for $\beta=0.5$ ($\alpha\beta=1$).}
\end{figure*}

\begin{figure*}[!ht]
\begin{center}
\includegraphics*[width=1.0\textwidth]{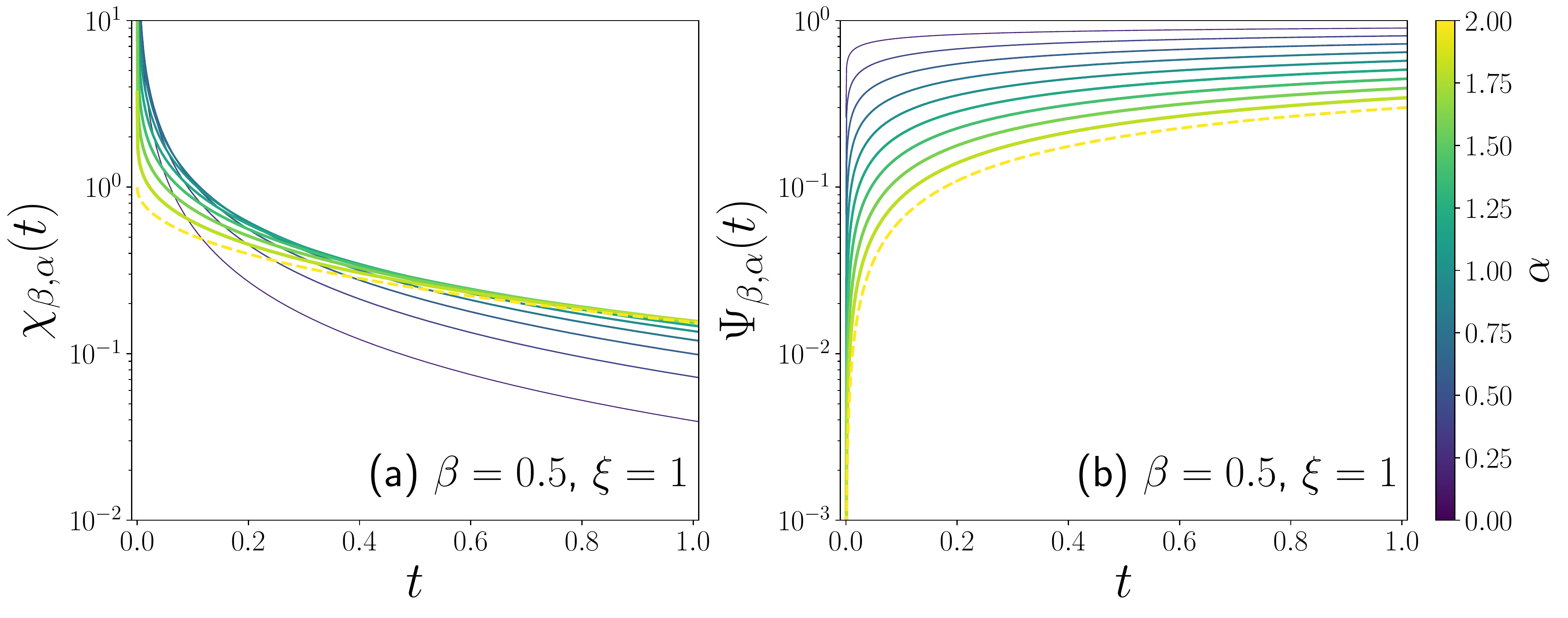}
\end{center}
\vspace{-5mm}
\caption{\label{Fig_2} (a) Jump density $\chi_{\beta,\alpha}(t)$ and (b) 
cumulated probability $\Psi_{\beta,\alpha}(t)$  as a function of $t$.
We explore the results for $\beta=0.5$  and different values  $0< \alpha \leq 2$ 
codified in the colorbar. The results were obtained numerically using $\xi=1$ 
and Eqs. (\ref{Laplainv}) and (\ref{cumulgenfract}). 
We represent with dashed lines the results for the particular case $\alpha\beta=1$.}
\end{figure*}

Beforehand we observe two limiting cases of `extreme' behaviors:
For infinitesimally small $\alpha$ we have $\lim_{\alpha\rightarrow 0+} {\tilde \chi}_{\beta,\alpha}(s) = 1$ and hence 
$\lim_{\alpha\rightarrow 0+}\chi_{\beta,\alpha}(t) = {\cal L}^{-1}\{1\} =\delta(t)$, i.e. the smaller
$\alpha$ the more the jump density becomes localized around $t=0+$ and taking for $\alpha\rightarrow 0+$
the shape of a $\delta(t)$-peak. In this limiting case the first arrival happens `immediately' 
when the observation starts at $t=0+$ with an extremely fast jump dynamics (Eqs. (\ref{remarkable-prop})).
On the other hand in the limit $\alpha\rightarrow \infty$ the Laplace transform of the jump density
$(1+\xi^{-1}s^{\beta})^{-\alpha}$ is equal to one at $s=0$ and nonzero extremely localized around
$s\approx 0$ dropping to zero immediately at $s\neq 0$. As a consequence
in this limit the jump density becomes extremely delocalized with quasi-equally 
distributed waiting times $\chi_{\beta,\alpha}(t) \sim \epsilon \rightarrow 0+$ for all $t >0$ thus extremely long waiting times may occur 
(shown subsequently by Eq. (\ref{tlarge-fat-tailed})).
In order to determine the time domain representation of (\ref{jump-gen-fractional-laplacetr}) it is convenient
to introduce the Pochhammer symbol which is defined as \cite{Mathai2010}
\begin{equation}
\label{Pochhammer}
(c)_m =  
\frac{\Gamma(c+m)}{\Gamma(c)} = \left\{\begin{array}{l} 1 ,\hspace{1cm} m=0 \\ \\ 
                                                            
     c(c+1)\ldots (c+m-1) ,\hspace{1cm} m=1,2,\ldots \end{array}\right.
\end{equation}
Despite $\Gamma(c)$ is singular at $c=0$ the Pochhammer symbol can be defined also
for $c=0$ by the limit $(0)_m= \lim_{c\rightarrow 0+}(c)_m=\delta_{m0}$ which is also fulfilled by the right hand side of (\ref{Pochhammer}). Then $(c)_m$ is defined
for all $c \in \mathbb{C}$.
The jump density defined by (\ref{jump-gen-fractional-laplacetr}) can be evaluated by accounting for
\begin{equation}
\label{simaliarlyasmittag-leffler}
\ds {\tilde \chi}_{\beta,\alpha}(s) = 
\xi^{\alpha} s^{-\beta\alpha }(1+\xi s^{-\beta})^{-\alpha}
=  \xi^{\alpha} 
\sum_{m=0}^{\infty} (-1)^m \frac{(\alpha)_m}{m!} \xi^m s^{-\beta (m+\alpha)}  
,\hspace{1cm}  \sigma=\Re(s) >\xi^{\frac{1}{\beta}}   .
\end{equation}
The time domain representation of the jump density of the GFPP is then obtained by using Laplace inversion
(\ref{Lplaceinversionnegativegamma}), namely (See also \cite{PolitoCahoy2013})
\begin{equation}
 \label{Laplainv}
 \begin{array}{l}
\ds  \chi_{\beta,\alpha}(t) = {\cal L}^{-1}\left\{  \frac{\xi^{\alpha}}{(s^{\beta}+\xi)^{\alpha}} \right\}=
 \frac{e^{\sigma t}}{2\pi} 
 \int_{-\infty}^{\infty} e^{i\omega t} \frac{\xi^{\alpha}}{(\xi + (\sigma +i\omega)^{\beta})^{\alpha}} {\rm d}\omega 
 \\ \\
 \ds \hspace{0.5cm} = \xi^{\alpha} \sum_{m=0}^{\infty} (-1)^m \frac{(\alpha)_m}{m!} 
 \xi^m {\cal L}^{-1}\{s^{-\beta (m+\alpha)}\} ,\hspace{0.25cm} t>0, \hspace{0.2cm} \sigma=\Re\{s\}>\xi^{\frac{1}{\beta}}, \hspace{0.2cm} 0<\beta\leq 1, 
 \hspace{0.2cm} \alpha >0\\ \\
 \ds \hspace{0.5cm}  = \xi^{\alpha} t^{\beta\alpha-1} \sum_{m=0}^{\infty}
 \frac{(\alpha)_m}{m!}\frac{(-\xi t^{\beta})^m}{\Gamma(\beta m+ \alpha\beta)} = 
 \xi^{\alpha} t^{\beta\alpha-1} E_{\beta,\alpha\beta}^{\alpha}(-\xi t^{\beta}) .
 \end{array} 
 \end{equation}
In expression (\ref{Laplainv}) appears a generalization of the Mittag-Leffler 
function which was introduced by Prabhakar \cite{Prabhakar1971} defined by
\begin{equation}
 \label{genmittag-Leff}
 E_{a,b}^c(z) = \sum_{m=0}^{\infty}
 \frac{(c)_m}{m!}\frac{z^m}{\Gamma(am + b)} ,\hspace{0.5cm} \Re\{a\} >0 ,
 \hspace{0.5cm} \Re\{b\} > 0 ,\hspace{0.5cm} c,\, z \in \mathbb{C}
\end{equation}
where with $(0)_m =\delta_{m0}$ we have $E_{a,b}^0(z) =1$. 
The generalized Mittag-Leffler function (\ref{genmittag-Leff}) was
analyzed by several authors \cite{Mathai2010,ShulkaPrajabati2007,HauboldMathaiSaxena2011,GaraGarrappa2018}.
This series converges absolutely in the entire complex $z$-plane.
For $t$ small the jump density behaves as
\begin{equation}
 \label{lowest-order-in-t}
 \lim_{t \rightarrow 0+} \chi_{\beta,\alpha}(t)= \lim_{t \rightarrow 0} \frac{\xi^{\alpha}}{\Gamma(\alpha\beta)} t^{\alpha\beta-1} 
\rightarrow \left\{\begin{array}{l} 0    ,\hspace{1cm}    \alpha\beta>1, \\ \\
 \xi^{\frac{1}{\beta}} , \hspace{1cm} \alpha\beta=1, \\ \\
         \infty \hspace{1cm}   \alpha\beta <1.
        \end{array} \right.
\end{equation}
The power-law asymptotic behavior $\sim t^{\alpha\beta-1}$ occurring for $t$ small 
is different from the fractional Poisson process 
which has $\sim t^{\beta-1}$ and is reproduced for $\alpha=1$ 
(See the order $n=0$ in the Mittag-Leffler density
(\ref{Mittag-Leffler-waiting-time-PDF})). The GFPP jump density allows exponents $\alpha\beta >1$ where the jump density approaches zero for $t\rightarrow 0+$.
This behavior does not exist in the case $\alpha=1$ of the fractional Poisson process.
\\[2mm]
We see in the Figures \ref{Fig_1} and \ref{Fig_2} that
$\lim_{t\rightarrow 0+}\chi_{\beta,\alpha}(t) \sim t^{\alpha\beta-1} \rightarrow +\infty$ occurs
in the regime $\alpha\beta <1$. The jump density in this regime is extremely high at short observation times, 
where this tendency is amplified for $\alpha\rightarrow 0+$ with $\delta(t)$-peak localized jump densities.
Further we observe in Figure \ref{Fig_1} the `phase transition' at $\alpha\beta=1$ separating the behaviors of Eq. (\ref{lowest-order-in-t})
for $t\rightarrow 0+$.
On the other hand for $\alpha\rightarrow \infty$ the jump density exhibits an extremely broad distribution of waiting times.
For $\alpha=1$ the density (\ref{Laplainv}) recovers the Mittag-Leffler
type jump density of the fractional Poisson process (\ref{Mittag-Leffler-waiting-time-PDF}), 
and for $\beta=1$, $\alpha>0$ is an Erlang type density where non-integer $\alpha >0$ are allowed 
(See Eq. (\ref{beta1alphachi})). For $\alpha=1$, 
$\beta=1$ the jump density $\chi_{1,1}(t))=\xi e^{-\xi t}$ of the standard Poisson process is reproduced.
For $\alpha=1$ the expression of the jump density (\ref{Laplainv}) recovers the jump density (\ref{Mittag-Leffler-waiting-time-PDF})
of the fractional Poisson process. Integrating (\ref{Laplainv}) yields the cumulated distribution
(\ref{waitintimedis}) of first arrivals 
\begin{equation}
\label{cumulgenfract}
\ds \Psi_{\beta,\alpha}(t)= \int_0^{t}  \chi_{\beta,\alpha}(\tau){\rm d}\tau  = 
 \xi^{\alpha} t^{\alpha\beta} \sum_{m=0}^{\infty}\frac{(\alpha)_m}{m!} 
 \frac{(-\xi t^{\beta})^m}{\Gamma(\beta m + \alpha\beta+1)} =  
 \xi^{\alpha} t^{\alpha\beta} E_{\beta,(\alpha\beta+1)}^{\alpha}(-\xi t^{\beta}) .
\end{equation}
The probability distribution (\ref{cumulgenfract}) is plotted for different values of
$0<\beta<1$, and $\alpha=2$ in Figure \ref{Fig_1} and for different values $\alpha >0$ in Figure \ref{Fig_2} for $\beta=0.5$.
For small $\xi^{\frac{1}{\beta}}t$, this distribution is dominated by order
$n=0$, namely
\begin{equation}
 \label{smalldmt}
 \Psi_{\beta,\alpha}(t) 
 \approx \frac{\xi^{\alpha}}{\Gamma(\alpha\beta+1)} t^{\alpha\beta} ,\hspace{1cm} 
 \xi^{\frac{1}{\beta}}t \rightarrow 0 .
\end{equation}
We see in this relation that in the admissible range $0<\beta\leq 1$, $\alpha>0$ the initial 
condition $\Psi_{\beta,\alpha}(0)=0$ is always fulfilled.
We hence observe the following distributional properties  
\begin{equation}
 \label{remarkable-prop}
 \begin{array}{l}
 \ds \lim_{\alpha\rightarrow 0+} \Psi_{\beta,\alpha}(t) = \lim_{\alpha\rightarrow 0+} \frac{\xi^{\alpha} t^{\alpha\beta} }{\Gamma(\alpha\beta+1)} =\Theta(t)  ,\\ \\
 \ds \lim_{\alpha\rightarrow 0+} \chi_{\beta,\alpha}(t) = \frac{d}{dt} \Theta(t) = \delta(t) 
 \end{array}
\end{equation}
and similar relations are obtained in the limit $\alpha\beta \rightarrow 0+$.
This behavior is consistent with the observation that in the Laplace domain ${\tilde \Psi}_{\beta,\alpha}(s)=s^{-1}{\tilde \chi}_{\beta,\alpha}(s)$ for $\alpha\rightarrow 0$ behaves as $s^{-1}$.
The $t^{\alpha\beta}$-power-law behavior can be observed in the Figures \ref{Fig_1} and \ref{Fig_2}.
The smaller $\alpha\beta \rightarrow 0$ the more (\ref{smalldmt}) approaches a Heaviside step
function shape where for small observation times an extremely 
high jump frequency occurs so that $\Psi_{\beta,\alpha}(t)$ rapidly approaches maximum value one.
The survival probability then is given by
\begin{equation}
 \label{survivalgenfractpro}
 \Phi_{\beta,\alpha}^{(0)}(t)= 
 \int_t^{\infty}\chi_{\beta,\alpha}(\tau){\rm d}\tau  =1-\Psi_{\beta,\alpha}(t)= 
 1- \xi^{\alpha} t^{\alpha\beta} E_{\beta,(\alpha\beta+1)}^{\alpha}(-\xi t^{\beta})
\end{equation}
with $\lim_{t\rightarrow\infty} \Phi_{\beta,\alpha}^{(0)}(t)=0$. 
For $\alpha=1$ (\ref{survivalgenfractpro}) becomes 
\begin{equation}
 \label{fractPoisson}
 \begin{array}{l}
 \ds \Phi_{\beta,1}^{(0)}(t) = 1- \xi t^{\beta} E_{\beta,(\beta+1)}^{1}(-\xi t^{\beta}) \\ \\ \ds  \hspace{0.5cm} =
 1+t^{\beta}\sum_{m=0}^{\infty}(-1)^{m+1}\xi^{m+1}\frac{(-\xi t^{\beta})^m}{\Gamma(\beta m+\beta+1)}
 = \sum_{m=0}^{\infty} \frac{(-\xi t^{\beta})^m}{\Gamma(\beta m+1)} = E_{\beta}(-\xi t^{\beta})
 \end{array}
\end{equation}
reproducing the Mittag-Leffler survival probability (\ref{survival}) of the fractional Poisson process
Mittag-Leffler jump density
\begin{equation}
 \label{jumpdensityyfrac}
 \chi_{\beta,1}(t) = 
 \chi_{\beta}(t)= \xi t^{\beta-1}E_{\beta,\beta}(-\xi t^{\beta}) = 
 \frac{d}{dt}(1-E_{\beta}(-\xi t^{\beta}))
\end{equation}
coinciding with the known results \cite{Laskin2003}.
On the other hand let us consider $\beta=1$ and $\alpha >0$ which gives
\begin{equation}
 \label{beta1alphachi}
 \chi_{1,\alpha}(t) = \xi^{\alpha} t^{\alpha-1}E^{\alpha}_{1,\alpha}(-\xi t) = 
 \xi^{\alpha}\frac{t^{\alpha-1}}{\Gamma(\alpha)} \sum_{m=0}^{\infty}\frac{(-\xi t)^m}{m!} = 
 \xi^{\alpha}\frac{t^{\alpha-1}}{\Gamma(\alpha)} e^{-\xi t} 
\end{equation}
i.e. for $\beta=1$ the jump density is light-tailed with exponentially 
evanescent behavior for $t\rightarrow \infty$, but for $t\rightarrow 0$ 
we get power-law behavior $\chi_{1,\alpha}(t) \sim t^{\alpha-1}$.
The density (\ref{beta1alphachi}) is the so-called Erlang density 
and the renewal process generated with this density is referred to as Erlang process
\cite{GorenfloMainardi2013}.
For $\alpha=1$ the Erlang density (\ref{beta1alphachi}) again recovers the density $\chi_{1,1}(t)= \xi e^{-\xi t}$ of the 
standard Poisson process.
\\[2mm]

Let us now consider the behavior for large times $t\rightarrow \infty$.
Expanding (\ref{jump-gen-fractional-laplacetr}) for $|s|$ small yields\footnote{The large time behavior 
emerges for $\xi\rightarrow\infty$ 
(or equivalently for small $|s|\rightarrow 0$) in (\ref{jump-gen-fractional-laplacetr}).}
\begin{equation}
 \label{jump-gen-fractional}
 {\tilde \chi}_{\beta,\alpha}(s) = \left(1+\frac{s^{\beta}}{\xi}\right)^{-\alpha} = \sum_{m=0}^{\infty}
 \frac{(\alpha)_m}{m!}(-1)^m \xi^{(-m)}s^{m\beta} = 1-\frac{\alpha}{\xi} s^{\beta}+\ldots 
 \end{equation}
which yields for large observation times for $0<\beta<1$, $\alpha>0$ fat-tailed behavior\footnote{Where
$-\Gamma(-\beta)=\beta^{-1}\Gamma(1-\beta) >0$.}
 \begin{equation}
  \label{tlarge-fat-tailed}
  \chi_{\beta,\alpha}(t) \approx -\frac{\alpha}{\xi \Gamma(-\beta)}t^{-\beta-1} ,
  \hspace{0.5cm} 0<\beta< 1 ,\hspace{0.5cm} \alpha>0 ,\hspace{0.5cm}
  t\rightarrow\infty
 \end{equation}
where the exponent does not depend on $\alpha$.

\section{Generalized fractional Poisson distribution}
\label{GeneralizedFractPoissonDistri}

In this section we derive the generalized counterpart $\Phi^{(n)}_{\beta,\alpha}(t)$ to Laskin's fractional Poisson distribution (\ref{fractionalpoission-distribution}), i.e.
the probabilities for $n$ arrivals\footnote{See Eq. (\ref{nstepprobabilityuptot}).} 
within $[0,t]$ in a GFPP.
For the evaluation we
utilize the general relation (\ref{laplansteps}) 
together with (\ref{jump-gen-fractional-laplacetr}), namely
\begin{equation}
\label{generalizeedFractPoissonDis}
\Phi^{(n)}_{\beta,\alpha}(t) ={\cal L}^{-1}\left\{ 
\frac{1}{s}\left({\tilde \chi}^n_{\beta,\alpha}(s)-{\tilde \chi}^{n+1}_{\beta,\alpha}(s)\right)\right\}= 
{\cal L}^{-1}\left\{ 
\frac{1}{s}\left({\tilde \chi}_{\beta,n\alpha}(s)-{\tilde \chi}_{\beta,(n+1)\alpha}(s)\right) \right\}
,\hspace{1cm} n=0,1,2,\ldots 
\end{equation}
where ${\tilde \chi}^n_{\beta,\alpha}(s) =\frac{\xi^{n\alpha}}{(\xi+s^{\beta})^{n\alpha}}$. We then obtain by
using Eq. (\ref{Laplainv})

\begin{equation}
 \label{chnprim}
 \begin{array}{l}
 \Psi_{\beta,n\alpha}(t) = \ds {\cal L}^{-1}\left\{\frac{1}{s}{\tilde \chi}_{\beta,n\alpha}(s)\right\} = 
 \xi^{n\alpha} \sum_{m=0}^{\infty} (-1)^m \frac{(n\alpha)_m}{m!} 
 \xi^m {\cal L}^{-1}\{s^{-\beta m- n\alpha\beta -1)}\} ,\hspace{1cm} \sigma=\Re\{s\} > \xi^{\frac{1}{\beta}} \\ \\
 \ds \hspace{0.5cm} =  \xi^{n\alpha} \sum_{m=0}^{\infty}  \frac{(n\alpha)_m}{m!} 
 \frac{(-1)^m \xi^m t^{m\beta+n\alpha\beta}}{\Gamma(m\beta+n\alpha\beta+1)} = 
 \xi^{n\alpha} t^{n\alpha\beta} E^{n\alpha}_{\beta,(n\alpha\beta+1)}(-\xi t^{\beta}) ,  \hspace{0.5cm} n=0,1,2,\ldots 
 \end{array}
\end{equation}
%
%
%
With this result, from (\ref{generalizeedFractPoissonDis}) we obtain the probability of {\it exactly} $n$ arrivals within $[0,t]$ as
\begin{equation}
 \label{Generalized-Fractional-Poisson-Distribution}
 \begin{array}{l}
   \Phi^{(n)}_{\beta,\alpha}(t)  = 
   \xi^{n\alpha} t^{n\alpha\beta} \left( E^{n\alpha}_{\beta,(n\alpha\beta+1)}(-\xi t^{\beta}) 
 - \xi^{\alpha} t^{\alpha\beta} E^{(n+1)\alpha}_{\beta,((n+1)\alpha\beta+1)}(-\xi t^{\beta}) \right) ,
  \\ \\
 \hspace{0.5cm} = \Psi_{\beta,n\alpha}(t)-\Psi_{\beta,(n+1)\alpha}(t)  ,\hspace{0.5cm} 0<\beta\leq 1 ,\hspace{0.2cm} \alpha >0 ,\hspace{0.2cm} n=0,1,2,\dots 
 \end{array}
\end{equation}
where $E^c_{a,b}(z)$ denotes the generalized Mittag-Leffler function defined in Eq. (\ref{genmittag-Leff}). This expression was also obtained in Ref. \cite{PolitoCahoy2013} (Eq. (2.8) in that paper).
We call the distribution $\Phi^{(n)}_{\beta,\alpha}(t)$ in Eq. (\ref{Generalized-Fractional-Poisson-Distribution}) the {\it generalized fractional Poisson distribution} (GFPD). 
In the second line of Eq. (\ref{Generalized-Fractional-Poisson-Distribution}) we have taken into account that cumulated probabilities $\Psi_{\beta,\nu \alpha}(t) =\int_0^t\chi_{\beta,\nu\alpha}(t){\rm d}t$ indicate
the probability for {\it at least} $\nu$ arrivals within $[0,t]$. The  $\Psi_{\beta,\nu \alpha}(t)$ are 
determined by Eq. (\ref{cumulgenfract}) by replacing there $\alpha\rightarrow \nu\alpha$.
The case $n=0$ is also covered in relation (\ref{Generalized-Fractional-Poisson-Distribution}) and yields 
$\Phi^{(0)}_{\beta,\alpha}(t)=1-\Psi_{\beta,\alpha}(t)$ of Eq. (\ref{survivalgenfractpro}). 
We can then write distribution
(\ref{Generalized-Fractional-Poisson-Distribution}) in the following compact form
\begin{equation}
 \label{Gen-Frac-Poisson}
 \Phi^{(n)}_{\beta,\alpha}(t) = (\xi t^{\beta})^{n\alpha} \left\{\frac{1}{\Gamma(n\alpha\beta +1)}+
 \sum_{m=1}^{\infty} A_{\alpha,\beta}^{m,n}(t^{\beta}\xi) \, (-\xi t^{\beta})^m \right\} 
\end{equation}
with 
\begin{equation}
 \label{coefficient-time-dep}
 A_{\alpha,\beta}^{m,n}(t^{\beta}\xi) = \frac{(n\alpha)_m}{m!\Gamma(n\alpha\beta+\beta m+1)}+
 \frac{(\xi t^{\beta})^{\alpha-1}((n+1)\alpha)_{m-1}}
 {(m-1)!\Gamma(n\alpha\beta+\beta m +(\alpha-1)\beta +1)} 
\end{equation}
for $m=1,2,\ldots$ and $n=0,1,2,\ldots$. The GFPD is a dimensionless probability distribution 
depending only on the dimensionless time $t \xi^{\frac{1}{\beta}} $
where $\xi^{-\frac{1}{\beta}}$ has dimension of time and defines a characteristic time scale in
the GFPP.
It follows that for small (dimensionless) times the GFPD behaves as
\begin{equation}
 \label{limitsmalltilmes}
 \Phi^{(n)}_{\beta,\alpha}(t) \approx \frac{(\xi t^{\beta})^{n\alpha}} {\Gamma(n\alpha\beta +1)} ,\hspace{1cm} 
 t\xi^{\frac{1}{\beta}} \rightarrow 0 ,\hspace{1cm} n=0,1,2,\ldots
\end{equation}
which is the inverse Laplace transform of $\xi^{n\alpha}s^{-n\alpha\beta -1}$, 
i.e. of the order $m=0$ in the expansion (\ref{chnprim}). It follows that the general relation
\begin{equation}
 \label{initialcon}
 \Phi^{(n)}_{\beta,\alpha}(t)\Big|_{t=0} = \delta_{n0}
\end{equation}
is fulfilled, i.e. at $t=0$ per construction the walker is on his departure site.
Further of interest is the asymptotic behavior for large (dimensionless) times $t\xi^{\frac{1}{\beta}}$.
To this end, let us expand the Laplace transform for small $s\rightarrow 0$ 
in (\ref{generalizeedFractPoissonDis}) up to the 
lowest non-vanishing order in $\frac{s^{\beta}}{\xi}$ to arrive at

\begin{equation}
\label{largetimes}
\Phi^{(n)}_{\beta,\alpha}(t) \approx \frac{\alpha}{\xi} {\cal L}^{-1}\{s^{\beta-1}\}= \frac{\alpha}{\Gamma(1-\beta)}
\left(t \xi^{\frac{1}{\beta}}\right)^{-\beta} ,\hspace{0.5cm} t\xi^{\frac{1}{\beta}} \rightarrow \infty 
,\hspace{0.5cm} 0<\beta<1, 
\hspace{0.5cm} \alpha >0 , \hspace{0.5cm} n=0 , 1, \dots ,\infty
\end{equation}
where this inverse power law holds universally for all $\alpha>0$ for $0< \beta < 1$ 
and is independent of the number of arrivals $n$ and contains the Laskin case $\alpha=1$. 
That is all `states' $n$ exhibit for $t\xi^{\frac{1}{\beta}}$ the same (universal) $t^{-\beta}$ 
inverse power-law decay. We conjecture that the asymptotic equal-distribution (\ref{largetimes})
in a wider sense can be interpreted as quasi-ergodic property of the GFPP.
\\[2mm]
Now let us consider the important limit $\alpha=1$ for the GFPD in more details. 
Then the functions (\ref{coefficient-time-dep}) become time independent coefficients, namely
\begin{equation}
 \label{coefficientsalpsha1}
  A_{1,\beta}^{m,n} = \frac{(n+m-1)!}{\Gamma(\beta(m+n)+1)}\left(\frac{1}{m!(n-1)!}+\frac{1}{(m-1)!n!}\right) = \frac{(n+m)!}{n!m!}
  \frac{1}{\Gamma(\beta(m+n)+1)} 
\end{equation}
and for the oder $m=0$ we have $\frac{1}{\Gamma(n\beta+1)}= \left(\frac{(n+m)!}{n!m!}
  \frac{1}{\Gamma(\beta(m+n)+1)}\right)\big|_{m=0} $ thus we get for the GFPP (\ref{Gen-Frac-Poisson}) for $\alpha=1$
  the expression
\begin{equation}
 \label{Laskin-case}
 \Phi^{(n)}_{\beta,1}(t) = \frac{(\xi t^{\beta})^n}{n!}\sum_{m=0}^{\infty}\frac{(n+m)!}{m!}
  \frac{(-\xi t^{\beta})^m}{\Gamma(\beta(m+n)+1)} 
\end{equation}
which we identify with Laskin's fractional Poisson distribution of Eq. (\ref{fractionalpoission-distribution}) 
\cite{Laskin2003,Laskin2009}.

\section{Expected number of arrivals in a GFPP}
\label{GenFractionalPoissonProcess}

Here we analyze the asymptotic behavior of the average number of arrivals ${\bar n}(t)$ within the
interval of observation $[0,t]$.
To this end we consider the generating function 
\begin{equation}
 \label{generatinggenfractalfunction}
 {\tilde {\cal G}}_{\beta,\alpha}(v,s) = \sum_{n=0}^{\infty} v^n {\tilde \Phi}^{(n)}_{\beta,\alpha}(s) = 
 \frac{1-{\tilde \chi}_{\beta,\alpha}(s)}{s(1-v {\tilde \chi}_{\beta,\alpha}(s))} =
 \frac{1}{s} \frac{(1+\frac{s^{\beta}}{\xi})^{\alpha}-1}{(1+\frac{s^{\beta}}{\xi})^{\alpha}-v} ,\hspace{1cm} |v|\leq 1
\end{equation}
with 
\begin{equation}
 \label{msteparrivale}
 {\tilde \Phi}^{(n)}_{\beta,\alpha}(s)= \frac{1-{\tilde \chi}_{\beta,\alpha}(s)}{s} {\tilde \chi}_{\beta,(n\alpha)}(s).
\end{equation}
The expected number ${\bar n}_{\beta,\alpha}(t)$ the walker makes in
$[0,t]$ is obtained from 
${\bar n}_{\beta,\alpha}(t)= \frac{d}{dv}{\cal G}_{\beta,\alpha}(v,t)|_{v=1}= 
\sum_{n=0}^{\infty}n\Phi^{(n)}_{\beta,\alpha}(t)$ with the Laplace transform \cite{PolitoCahoy2013}
\begin{equation}
 \label{Laplacetranfostepsaverge}
 {\tilde {\bar n}}_{\beta,\alpha}(s) = 
 \frac{d}{dv} {\tilde {\cal G}}_{\beta,\alpha}(v,s)\Big|_{v=1}= 
 \frac{{\tilde \chi}_{\beta,\alpha}(s)}{s (1-{\tilde \chi}_{\beta,\alpha}(s))} =
 \frac{1}{s[(\frac{s^{\beta}}{\xi}+1)^{\alpha}-1]} .
\end{equation}
The Laplace transform behaves as 
${\tilde {\bar n}}_{\beta,\alpha}(s) \approx \frac{\xi}{\alpha}s^{-\beta-1} $ for $|s|\rightarrow 0 $. 
We obtain hence the asymptotic behavior for 
$t\xi^{\frac{1}{\beta}} \rightarrow \infty$ large
\begin{equation}
 \label{largetimesteps}
 {\bar n}_{\beta,\alpha}(t) 
\approx \frac{\xi }{\alpha \Gamma(\beta+1)} t^{\beta} 
,\hspace{1cm} 0<\beta \leq 1 ,\hspace{1cm} \alpha >0, \hspace{1cm} t\xi^{\frac{1}{\beta}}\rightarrow\infty .
\end{equation}
This result includes the Erlang regime $\beta=1$ where for all $\alpha >0$ 
the average number of arrivals ${\bar n}_{1,\alpha}(t) \approx \frac{\xi}{\alpha} t $ 
increases linearly for large $t\xi^{\frac{1}{\beta}}$ and $\alpha=1$ recovers the well-known 
result of the standard Poisson process, e.g. \cite{GorenfloMainardi2013}.
On the other hand, we obtain the asymptotic behavior for $t\xi^{\frac{1}{\beta}} \rightarrow 0$ 
small when we expand ${\tilde {\bar n}}_{\beta,\alpha}(s)$ with for $|s|\rightarrow \infty$
and obtain
\begin{equation}
 \label{deductionoftsmall}
 {\bar n}_{\beta,\alpha}(t) = {\cal L}^{-1}\left\{\frac{1}{s[(\frac{s^{\beta}}{\xi}+1)^{\alpha}-1]}\right\} \approx {\cal L}^{-1}\left\{\xi^{\alpha}s^{-\alpha\beta-1}\right\}=
 \frac{(\xi t^{\beta})^{\alpha} }{\Gamma(\alpha\beta+1)} ,\hspace{1cm} t\xi^{\frac{1}{\beta}}\rightarrow 0.
\end{equation}
It is important to note that both asymptotic expressions (\ref{largetimesteps}) and (\ref{deductionoftsmall}) 
coincide for $\alpha=1$
recovering the known expression of the fractional Poisson process.

\section{Montroll-Weiss CTRW on undirected networks}
\label{CTRWnetworks}

As an application we consider a random walk on an undirected network subordinated 
to a GFPP and derive the generalization of the fractional Kolmogorov-Feller equation.
Then in subsequent Section \ref{definitionGFPP} we develop this process for the
infinite $d$-dimensional integer lattice $\mathbb{Z}^d$
and analyze the resulting `generalized fractional diffusion'. 
To this end let us briefly outline the Montroll-Weiss CTRW on undirected networks.
We assume an undirected connected network with $N$ nodes which we 
denote with $p=1,\ldots ,N$. For our convenience we employ Dirac's $|bra\rangle\langle ket|$ notation.
In an undirected network the positive-semidefinite $N \times N$
Laplacian matrix ${\mathbf L} =(L_{pq})$ characterizes
the network topology and is defined by \cite{TMM-APR-ISTE2019,NohRieger2004,RiascosMateos2014} 
\begin{equation}
 \label{Laplacianmat}
 L_{pq}=K_p \delta_{pq}-A_{pq}
\end{equation}
where ${\mathbf A} = (A_{pq})$ denotes the adjacency matrix having elements one 
if a pair of nodes is connected and zero otherwise. Further we forbid that nodes are connected with themselves
thus $A_{pp}=0$.
In an undirected network adjacency and Laplacian matrices are symmetric. The diagonal 
elements $L_{pp}=K_p$ of the Laplacian matrix are referred to as the degrees 
of the nodes $p$ counting the number of neighbor nodes of a node $p$ with $K_p=\sum_{q=1}^NA_{pq}$. 
The one-step transition matrix ${\mathbf W} =(W_{pq})$ relating the 
network topology 
with the random walk then is defined by \cite{TMM-APR-ISTE2019,NohRieger2004}
\begin{equation}
 \label{one-step}
  W_{pq} =\frac{1}{K_p}A_{pq}= \delta_{pq}-\frac{1}{K_p}L_{pq} 
\end{equation}
which generally is a non-symmetric matrix for networks with variable degrees
$K_i\neq K_j$ $ (i\neq j)$.
The one-step transition matrix $W_{pq}$ defines the conditional probability that 
the walker which is on node $p$ jumps in one step to node $q$ where in one step only neighbor nodes
with equal probability $\frac{1}{K_p}$ can be reached.
\\[2mm]
We consider now a random walker that performs a CTRW with IID random steps at
random times $0\leq t_1,t_2,\ldots t_n,\ldots \infty $ on the network where the observation starts at $t=0$. 
Each step of the walker from one to another node is associated with a jump event or arrival in a CTRW
with identical transition probability $(W_{pq})$ for a step from node $p$ to node $q$.
We introduce the {\it transition matrix} ${\mathbf P}(t)= (P_{ij}(t))$ 
indicating the {\it probability} to find the walker at time $t$ on node $j$
under the condition that the walker was sitting at node $i$ at time $t=0$ when the observation starts.

The transition matrix fulfills the normalization condition $\sum_{j=1}^NP_{ij}(t) =1$ 
and stochasticity of the transition matrix implies that $0\leq P_{ij}(t) \leq 1$.
We restrict us on connected undirected networks and allow variable degrees. 
In such networks the transition matrix is non-symmetric 
$P_{ij}(t) \neq P_{ji}(t)$ (See \cite{TMM-APR-ISTE2019} for a detailed analysis).
Assuming the initial condition
${\mathbf P}(t=0)$, then the probability to find the walker
on node $j$ at time $t$ is given by a series of the form \cite{Gorenflo2010,GorenfloMainardi2013,Cox1967}
\begin{equation}
 \label{the-CTRW}
 {\mathbf P}(t) =  {\mathbf P}(0) \sum_{n=0}^{\infty} \Phi^{(n)}(t) {\mathbf W}^n 
\end{equation}
where the $\Phi^{(n)}(t)$ indicate the probabilities of $n$ arrivals, 
i.e. that the walker made $n$ steps within interval $[0,t]$ of relation (\ref{nstepprobabilityuptot}). 
It is straight-forward to see in (\ref{the-CTRW}) 
that the normalization condition $\sum_{j=1}^NP_{ij}(t)=\sum_{n=0}^{\infty} \Phi^{(n)}(t) =1$ is fulfilled 
during the entire observation time $t\geq 0$.
The convergence of this series is proved by using that ${\mathbf W}$ has uniquely eigenvalues 
$|\lambda_m|\leq 1$ \cite{TMM-APR-ISTE2019} together with $|{\tilde \chi}(s)| \leq 1$ for the 
Laplace transforms of the jump density.
Let us assume that at $t=0$ the walker 
is sitting on its departure node $P_{ij}(0)=(\delta_{ij})$, then the Laplace transform of (\ref{the-CTRW}) writes
\begin{equation}
 \label{occupationproblaplace}
 {\tilde {\mathbf P}}(s) = \frac{\left(1-{\tilde \chi}(s)\right)}{s} \left\{{\mathbf 1}-
 {\tilde \chi}(s){\mathbf W}\right\}^{-1} .
\end{equation}
Let us take into account the canonic representation of the one-step 
transition matrix \cite{TMM-APR-ISTE2019}
\begin{equation}
 \label{canonic-one-step}
 {\mathbf W} = |\Phi_1\rangle\langle{\bar \Phi}_1| 
 +\sum_{m=2}^N\lambda_m |\Phi_m\rangle\langle{\bar \Phi}_m| 
\end{equation}
where we have always a unique eigenvalue $\lambda_1=1$\footnote{reflecting row-normalization
$\sum_{q=1}^N W_{pq}=1$. } $|\lambda_m| < 1$ ($m=2,\ldots ,N$) 
\cite{TMM-APR-ISTE2019}\footnote{In this demonstration we ignore the cases 
of bipartite graphs where a unique eigenvalue $-1$ occurs.}. 
In Eq. (\ref{canonic-one-step}) we have introduced the right- and left 
right eigenvectors $|\Phi_j\rangle$ and $\langle{\bar \Phi}_j|$ of ${\mathbf W}$, 
respectively. The first term that corresponds to $\lambda_1=1$ in 
$|\Phi_1\rangle\langle{\bar \Phi}_1|$ indicates the stationary distribution. 
We can then write the canonic representation of (\ref{occupationproblaplace}) in the form
\begin{equation}
 \label{canonic-occupation-probabilities}
 {\tilde {\mathbf P}}(s) = \sum_{m=1}^N {\tilde P}(m,s) 
 |\Phi_m\rangle\langle{\bar \Phi}_m| 
\end{equation}
where ${\tilde {\mathbf P}}(s)$ has the eigenvalues
\begin{equation}
 \label{Montroll-Weiss}
{\tilde P}(m,s) = \frac{\left(1-{\tilde \chi}(s)\right)}{s}\frac{1}{(1- \lambda_m{\tilde \chi}(s))}.
\end{equation}
This expression is the celebrated Montroll-Weiss formula \cite{MontrollWeiss1965}. Since $\lambda_1=1$ we see that
${\tilde P}(1,s) = \frac{1}{s}$ thus the stationary amplitude always is of the form 
$P(1,t) ={\cal L}^{-1}\{\frac{1}{s}\}= \Theta(t)$.

\section{Generalized fractional Kolmogorov-Feller equation}
\label{definitionGFPP}
With the general remarks on CTRWs of the previous section 
we now analyze a random walk on an undirected network which is
subordinated to a GFPP. For this diffusional process we derive a generalization of the fractional Kolmogorov-Feller equation 
\cite{SaichevZaslavski1997,Laskin2003}.
To this end let us first consider the
Laplace transforms of the GFPD $\Phi^{(n)}_{\beta,\alpha}(t)$ 
(See Eq. (\ref{generalizeedFractPoissonDis}))
\begin{equation}
 \label{alplacephin}
 {\tilde \Phi}^{(n)}_{\beta,\alpha}(s) = 
 {\tilde \Phi}^{(0)}_{\beta,\alpha}(s)\frac{\xi^{n\alpha}}{(s^{\beta}+\xi)^{n\alpha}} ,
 \hspace{0.5cm} \alpha >0 ,\hspace{0.5cm} 0< \beta\leq 1 ,\hspace{0.5cm} n=0,1,2,\dots
\end{equation}
where 
\begin{equation}
 \label{njump-density}
 {\tilde \chi}_{\beta,\alpha}^{(n)}(s) = {\tilde \chi}_{\beta,(n\alpha)}(s)=  \frac{\xi^{n\alpha}}{(s^{\beta}+\xi)^{n\alpha}}
\end{equation}
and especially for $n=0$ we have
\begin{equation}
 \label{Phizeroalhbet}
 {\tilde \Phi}^{(0)}_{\beta,\alpha}(s) =\frac{1-{\tilde \chi}_{\beta,\alpha}(s)}{s}   =
 \frac{(s^{\beta}+\xi)^{\alpha}-\xi^{\alpha}}{s(s^{\beta}+\xi)^{\alpha}} .
\end{equation}
These Laplace transforms fulfill 
\begin{equation}
 \label{generalrelationngen}
 (s^{\beta}+\xi)^{\alpha} {\tilde \Phi}^{(n)}_{\beta,\alpha}(s) = \xi^{\alpha} {\tilde \Phi}^{(n-1)}_{\beta,\alpha}(s) ,\hspace{1cm} n=1,2,\ldots ,
\end{equation}
and for $n=0$ we have
\begin{equation}
 \label{Phizer-oalhbetnz-zero-case}
 (s^{\beta}+\xi)^{\alpha} {\tilde \Phi}^{(0)}_{\beta,\alpha}(s) = 
 \frac{(s^{\beta}+\xi)^{\alpha}-\xi^{\alpha}}{s} .
\end{equation}
In the time domain these equations write
\begin{equation}
\label{takeforms}
_0\!{\cal D}_t^{\beta,\alpha} \Phi^{(n)}_{\beta,\alpha}(t) =
\xi^{\alpha} \Phi^{(n-1)}_{\beta,\alpha}(t) ,\hspace{1cm} n=1,2,\ldots ,
\end{equation}
and
\begin{equation}
 \label{nzeroyields}
 _0\!{\cal D}_t^{\beta,\alpha}(\xi) \Phi^{(0)}_{\beta,\alpha}(t) = 
 {\cal L}^{-1}\left\{\frac{(s^{\beta}+\xi)^{\alpha}-\xi^{\alpha}}{s}\right\} 
 = K^{(0)}_{\beta,\alpha}(t) -\xi^{\alpha}\Theta(t)
\end{equation}
where 
$_0\!{\cal D}_t^{\beta,\alpha} \Phi(t)=: 
\int_0^t\Phi(\tau){\cal D}^{\beta,\alpha}(t-\tau){\rm d}\tau$ 
has to be read as a 
convolution. 
\\[2mm]
Now the goal is to derive the generalized counterpart to the fractional 
Kolmogorov-Feller equation \cite{SaichevZaslavski1997,Laskin2003}.
A major role is played by the kernel 
\begin{equation}
 \label{kernel-associated}
{\cal D}^{\beta,\alpha}(t) = {\cal L}^{-1} \left\{(s^{\beta}+\xi)^{\alpha}\right\} .
\end{equation}
This kernel has the explicit representation \cite{MichelitschRiascosGFPP2019} (and see also Appendix \ref{AppendLaplacetrafo})\footnote{Where we introduce $d^{\beta,\alpha}(t) = {\cal L}^{-1}\left\{s^{-\ceil{\alpha\beta}} (s^{\beta}+\xi)^{\alpha}\right\}$.}
\begin{equation}
 \label{en-res-mittag-leffl-gn-kernel}
 \begin{array}{l}
 \ds {\cal D}^{\beta,\alpha}(t) = {\cal L}^{-1} \left\{(s^{\beta}+\xi)^{\alpha}\right\} = \frac{d^{\ceil{\alpha\beta}}}{dt^{\ceil{\alpha\beta}}} \left\{\Theta(t) d^{\beta,\alpha}(t)\right\}  \\ \\ \ds 
 \hspace{0.5cm} = \left\{\begin{array}{l}
  \frac{d^{\ceil{\alpha\beta}}}{dt^{\ceil{\alpha\beta}}} 
  \left\{ \Theta(t) t^{\ceil{\alpha\beta}-\beta\alpha-1} 
  \sum_{m=0}^{\infty}\frac{\alpha!}{(\alpha-m)!m!} 
 \frac{(\xi  t^{\beta})^m}{\Gamma(\beta m +\ceil{\alpha\beta}-\beta\alpha)} \right\} ,
 \hspace{0.5cm}\alpha\beta \notin \mathbb{N} \\ \\
 \frac{d^{\alpha\beta}}{dt^{\alpha\beta}}\left\{
 \delta(t)+  \Theta(t) \frac{d}{dt}\sum_{m=0}^{\infty}\frac{\alpha !}{(\alpha-m)!m!} 
 \frac{(\xi t^{\beta})^m}{\Gamma(m\beta+1)} \right\}  ,
\hspace{1cm} \alpha\beta \in \mathbb{N} \end{array}\right.
 \\ \\ \ds 
 \hspace{0.5cm} = \left\{\begin{array}{l} \frac{d^{\ceil{\alpha\beta}}}{dt^{\ceil{\alpha\beta}}} 
 \left\{ \Theta(t) t^{\ceil{\alpha\beta}-\beta\alpha-1}
 E_{\alpha,\beta,(\ceil{\alpha\beta}-\alpha\beta)}(\xi t^{\beta}) \right\} ,\hspace{1cm} \alpha\beta \notin \mathbb{N}  \\ \\
\frac{d^{\alpha\beta}}{dt^{\alpha\beta}}\left(\delta(t) + 
\Theta(t)\frac{d}{dt}E_{\alpha,\beta,1}(\xi t^{\beta})\right) ,
\hspace{1cm} \alpha\beta \in \mathbb{N}. \end{array}\right. 
 \end{array}
\end{equation}
In this expression we introduced the {\it ceiling function} $\ceil{\gamma}$ indicating the smallest integer
greater or equal to $\gamma$.
In (\ref{en-res-mittag-leffl-gn-kernel}) occurs a generalized Mittag-Leffler type function
$E_{c,a,b}(z)$
which is related with the generalized Prabhakar Mittag-Leffler function (\ref{genmittag-Leff}) and given by
\begin{equation}
 \label{Mittag-Leffler-other-gen}
 E_{c,a,b}(z) = \sum_{m=0}^{\infty} \frac{c!}{(c-m)!m!} \frac{z^m}{\Gamma(a m+b)} =
 \sum_{m=0}^{\infty} \frac{(-c)_m}{m!} \frac{(-z)^m}{\Gamma(am+b)} = E_{a,b}^{-c}(-z)
 ,\hspace{0.5cm} 
  \Re\{a\} > 0 ,\hspace{0.5cm} \Re\{b\} > 0.
\end{equation}
The second kernel $K^{(0)}_{\beta,\alpha}(t)$ of equation (\ref{nzeroyields}) is then explicitly
obtained as \cite{MichelitschRiascosGFPP2019}
\begin{equation}
 \label{Kzeroexplicitform}
\ds  K^{(0)}_{\beta,\alpha}(t) ={\cal L}^{-1}\left\{\frac{(s^{\beta}+\xi)^{\alpha}}{s}\right\} = \left\{\begin{array}{l} \Theta(t)t^{-\alpha\beta} 
 E_{\alpha,\beta,1-\alpha\beta}(\xi t^{\beta}) , ,\hspace{1cm} 0<\alpha\beta <1 \\ \\
 \frac{d^{\ceil{\alpha\beta}-1}}{dt^{\ceil{\alpha\beta}-1}}
   \left\{\Theta(t) t^{\ceil{\alpha\beta}-\beta\alpha-1} 
 E_{\alpha,\beta,(\ceil{\alpha\beta}-\alpha\beta)}(\xi t^{\beta})\right\} 
 \hspace{0.5cm} \alpha\beta >1, \hspace{0.5cm} \alpha\beta \notin \mathbb{N} 
 \\ \\
 \frac{d^{\alpha\beta-1}}{dt^{\alpha\beta-1}}\left(\delta(t) + 
\Theta(t)\frac{d}{dt}E_{\alpha,\beta,1}(\xi t^{\beta})\right),
\hspace{1cm} \alpha\beta \geq 1 , \hspace{0.5cm} \alpha\beta \in \mathbb{N}. 
                                    \end{array}\right. 
\end{equation}
The transition probability matrix (\ref{the-CTRW})
of the GFPP walk then writes
\begin{equation}
\label{the-CTRW-GFPP}
 {\mathbf P}_{\beta,\alpha}(t) = {\mathbf P}(0) \Phi_{\beta,\alpha}^{(0)}(t)   + 
 {\mathbf P}(0) \sum_{n=1}^{\infty} \Phi_{\beta,\alpha}^{(n)}(t) {\mathbf W}^n .
\end{equation}
Then we obtain with relations (\ref{takeforms}), (\ref{nzeroyields}) the convolution equation\footnote{Read $ _0\!{\cal D}_t^{\beta,\alpha}\cdot {\mathbf P}_{\beta,\alpha}(t)=
 \int_0^t {\cal D}^{\beta,\alpha}(t-\tau){\mathbf P}_{\beta,\alpha}(\tau){\rm d}\tau =: \frac{d^{\ceil{\alpha\beta}}}{dt^{\ceil{\alpha\beta}}} \int_0^td^{\beta,\alpha}(t-\tau)
 {\mathbf P}_{\beta,\alpha}(\tau){\rm d}\tau$.} 
\begin{equation}
\label{generalizedfracKolmogorivFeller}
\begin{array}{l} 
\ds  _0\!{\cal D}_t^{\beta,\alpha}\cdot {\mathbf P}_{\beta,\alpha}(t)  = 
{\mathbf P}(0)\left( K^{(0)}_{\beta,\alpha}(t) -\xi^{\alpha}\Theta(t)\right)  +
\xi^{\alpha} {\mathbf P}_{\beta,\alpha}(t){\mathbf W}  ,\\ \\
\ds \int_0^t {\cal D}^{\beta,\alpha}(t-\tau){\mathbf P}_{\beta,\alpha}(\tau){\rm d}\tau =
{\mathbf P}(0)\left( K^{(0)}_{\beta,\alpha}(t) -\xi^{\alpha}\Theta(t)\right)  +
\xi^{\alpha} {\mathbf P}_{\beta,\alpha}(t){\mathbf W} .
\end{array}
 \end{equation}
We call equation (\ref{generalizedfracKolmogorivFeller}) 
the {\it generalized fractional Kolmogorov-Feller equation}. 
It has for $\alpha\beta \notin \mathbb{N}$ the explicit representation

\begin{equation}
 \label{explict-gen-fract}
 \begin{array}{l}
 \ds \frac{d^{\ceil{\alpha\beta}}}{dt^{\ceil{\alpha\beta}}} \int_0^t
 (t-\tau)^{\ceil{\alpha\beta}-\beta\alpha-1} 
 E_{\alpha,\beta,(\ceil{\alpha\beta}-\alpha\beta)}(\xi (t-\tau)^{\beta}) 
 {\mathbf P}_{\beta,\alpha}(\tau){\rm d}\tau  ,\hspace{0.5cm} \alpha\beta \notin \mathbb{N} \\ \\ \ds \hspace{0.5cm} = 
 {\mathbf P}(0) \left(K^{(0)}_{\beta,\alpha}(t)
 -\xi^{\alpha}\Theta(t)\right)  + 
 \xi^{\alpha} {\mathbf P}_{\beta,\alpha}(t){\mathbf W} ,\hspace{0.5cm} 0<\beta<1, \hspace{0.5cm} \alpha >0 .
 \end{array}
\end{equation}
Let us next consider the cases $\alpha=1$ of Laskin's fractional Poisson process,
and of $\alpha=1$, $\beta=1$ representing the standard Poisson process.
\\[2mm]
We first consider (\ref{explict-gen-fract}) in the fractional Poisson process limit 
$\alpha=1$ and $0<\beta<1$.
Then we get for the kernel (\ref{Kzeroexplicitform})
\begin{equation}
 \label{kernel0alphabet}
 K^{(0)}_{\beta,1}(t) = t^{-\beta}E_{1,\beta,1-\beta}(\xi t^{\beta})  = \Theta(t) \frac{t^{-\beta}}{\Gamma(1-\beta)} + \xi \Theta(t)
\end{equation}
$
$
with
\begin{equation}
\label{MLalpha1}
E_{1,\beta,1-\beta}(\xi t^{\beta}) = \frac{1}{\Gamma(1-\beta)} +\frac{\xi t^{\beta}}{\Gamma(1)} = 
\frac{1}{\Gamma(1-\beta)} +\xi t^{\beta} .
\end{equation}
Only the orders $m=0$ and $m=1$ in $\xi t^{\beta}$ are nonzero.
Then with $\ceil{\alpha\beta}=\ceil{\beta}=1$ we get for (\ref{explict-gen-fract})
\begin{equation}
 \label{first-step}
 \frac{d}{dt} \int_0^t (t-\tau)^{-\beta}E_{1,\beta,1-\beta}(\xi (t-\tau)^{\beta}) {\mathbf P}_{\beta,1}(\tau){\rm d}\tau  =
  {\mathbf P}(0) \left(K^{(0)}_{\beta,1}(t)
 -\xi \Theta(t)\right)  + 
 \xi {\mathbf P}_{\beta,1}(t){\mathbf W}
\end{equation}
which takes with (\ref{kernel0alphabet}) and (\ref{MLalpha1}) the form ($t>0$)
\begin{equation}
 \label{KFfractcasealpha1}
 \frac{d}{dt}\int_0^t (t-\tau)^{-\beta}\left(\frac{1}{\Gamma(1-\beta)} +\xi (t-\tau)^{\beta} \right){\mathbf P}_{\beta,1}(\tau){\rm d}\tau  = 
 \frac{t^{-\beta}}{\Gamma(1-\beta)} {\mathbf P}(0)  + 
 \xi {\mathbf P}_{\beta,1}(t){\mathbf W} .
\end{equation}
On the left hand side we identify $\frac{1}{\Gamma(1-\beta)}\frac{d}{dt}\int_0^t (t-\tau)^{-\beta}{\mathbf P}_{\beta,1}(\tau){\rm d}\tau = _0 \!D_t^{\beta}{\mathbf P}_{\beta,1}(t)$
with the Riemann-Liouville fractional derivative of order $\beta$. The second term on the left hand side yields 
$\xi \frac{d}{dt} \int_0^t{\mathbf P}_{\beta,1}(\tau){\rm d}\tau =\xi {\mathbf P}_{\beta,1}(t)$.
Hence we can write (\ref{KFfractcasealpha1}) in the form
\begin{equation}
 \label{fractKGFeq}
 _0 \!D_t^{\beta}{\mathbf P}_{\beta,1}(t) =   \frac{t^{-\beta}}{\Gamma(1-\beta)} {\mathbf P}(0)  -\xi {\mathbf P}_{\beta,1}(t) +
 \xi {\mathbf P}_{\beta,1}(t){\mathbf W} ,\hspace{1cm} 0<\beta< 1 .
\end{equation}
We identify this equation with the {\it fractional Kolmogorov-Feller equation} \cite{SaichevZaslavski1997,Laskin2003}. In this way we have proved that the 
{\it generalized fractional Kolmogorov-Feller equation} (\ref{generalizedfracKolmogorivFeller})
for $\alpha=1$ recovers the fractional Kolmogorov-Feller equation introduced in 
\cite{SaichevZaslavski1997} and see also the references
\cite{Gorenflo2010,GorenfloMainardi2013,MeerschaertEtal2011,HilferAnton1995}. 
%
%

In the fractional Kolmogorov-Feller equation (\ref{fractKGFeq}) occurs the `memory term' $\frac{t^{-\beta}}{\Gamma(1-\beta)} {\mathbf P}(0)$ which reflects the slow power-law decay of the influence of the initial condition. 
Generally for the GFPP the long-time memory (non-markovianity) is a consequence that the jump density is fat-tailed $\sim t^{-1-\beta}$, see Eq. (\ref{tlarge-fat-tailed}). Let us demonstrate briefly
that the same memory effect occurs in walks subordinated to GFPPs for any $\alpha>0$. To this end it is instructive to consider the generalized fractional Kolmogorov-Feller equation (\ref{generalizedfracKolmogorivFeller}) in the Laplace domain
\begin{equation}
 \label{laplacedomainkolofeller}
 (s^{\beta}+\xi)^{\alpha}{\tilde {\mathbf P}}_{\beta,\alpha}(s) = \frac{(s^{\beta}+\xi)^{\alpha}-\xi^{\alpha}}{s}{\mathbf P}(0)+\xi^{\alpha}{\mathbf W}{\tilde {\mathbf P}}_{\beta,\alpha}(s) .
\end{equation}
For $|s| \rightarrow 0$ small this equation takes the representation
\begin{equation}
 \label{smallsgenfrac}
 s^{\beta}{\tilde {\mathbf P}}_{\beta,\alpha}(s) =s^{\beta-1} {\mathbf P}(0) - \frac{\xi}{\alpha}{\tilde {\mathbf P}}_{\beta,\alpha}(s)+\frac{\xi}{\alpha}{\mathbf W}{\tilde {\mathbf P}}_{\beta,\alpha}(s) ,\hspace{1cm} 0<\beta<1 .
\end{equation}
Transforming Eq. (\ref{smallsgenfrac}) into the time domain yields fractional Kolmogorov-Feller equation (\ref{fractKGFeq}) 
as the {\it the asymptotic limit for $\xi^ {\frac{1}{\beta}}t \rightarrow \infty$} 
of the generalized fractional fractional Kolmogorov-Feller equation (\ref{generalizedfracKolmogorivFeller})
(with $\xi\rightarrow \xi' = \frac{\xi}{\alpha}$). It follows that the GFPP generates asymptotically the same memory effect as the fractional Poisson process with 
the memory term $\frac{t^{-\beta}}{\Gamma(1-\beta)} {\mathbf P}(0)$ ($0<\beta<1$) which is independent of $\alpha$.
The memory effect becomes especially pronounced when $\beta\rightarrow 0+$ with extremely slow decay. On the other hand in the limit $\beta\rightarrow 1-0$ the memory term takes $\delta(t)$-representation thus
the walk then becomes memoryless. 
\\[2mm]

Finally we consider the limit $\alpha=1$ and $\beta=1$ of standard Poisson. Then Eqs. (\ref{generalrelationngen})
and (\ref{Phizer-oalhbetnz-zero-case}) take in the time domain the form

\begin{equation}
 \label{causalbeta1n}
 \left(\frac{d}{dt}+\xi\right) \Phi_{1}^{(n)}(t) = \xi \Phi_{1}^{(n-1)}(t)
\end{equation}
and 
\begin{equation}
 \label{causalbeta1zero}
 \left(\frac{d}{dt}+\xi\right) \Phi_{1}^{(0)}(t) = \delta(t)
\end{equation}
where these relations define the standard Poisson process and are easily confirmed to be fulfilled by
the Poisson distribution
(\ref{standardpoisson}).
Plugging the last two equations into (\ref{generalizedfracKolmogorivFeller}) for $\beta=1$ and $\alpha=1$
yields relation
\begin{equation}
 \label{standardPoissonKolmogorovFeller}
 \frac{d}{dt} {\mathbf P}_{1,1}(t) = 
 {\mathbf P}(0) \delta(t) - \xi  {\mathbf P}_{1,1}(t) + \xi {\mathbf P}_{1,1}(t)  {\mathbf W} 
\end{equation}
which is known as the {\it Kolmogorov-Feller} equation of the standard Poisson process 
\cite{Laskin2003,Gorenflo2010,GorenfloMainardi2013}.
One can also recover the Kolmogorov Feller equation (\ref{standardPoissonKolmogorovFeller})
from Eq. (\ref{fractKGFeq}) in the limit $\beta \rightarrow 1-0$.

\section{Generalized fractional diffusion in $\mathbb{Z}^d$}
\label{genfractdiffusion}

In this section we analyze the features of a random walk subordinated to a GFPP 
in the infinite $d$-dimensional integer lattice $\mathbb{Z}^d$. We refer
the resulting diffusion process to as `generalized fractional diffusion'.
The eigenvalues of the one-step transition 
matrix (\ref{canonic-one-step}) for the $d$-dimensional infinite lattice
are given by the Fourier transforms \cite{TMM-APR-ISTE2019,TMM-APR-JPhys-A2017}
\begin{equation}
 \label{eigvals}
 \begin{array}{l}
\ds  \lambda(\vec{k})= 1-\frac{1}{2d}\mu(\vec{k}) ,\hspace{1cm} \mu(\vec{k}) = 2d-2\sum_{j=1}^d\cos(k_j) \\ \\
\ds  \lambda(\vec{k}) = \frac{1}{d}\sum_j^d\cos(k_j) ,\hspace{1cm} -\pi \leq k_j \leq \pi .
 \end{array}
\end{equation}
Here $\mu(\vec{k})$ denote the eigenvalues of the Laplacian matrix (\ref{Laplacianmat}) of the lattice. 
In this walk the walker in one jump can reach only connected neighbor nodes. Such walk refers 
to as `normal walk' \cite{NohRieger2004}. In the present section we subordinate a 
normal walk on the $\mathbb{Z}^d$ to a GFPP.
In this lattice the eigenvectors become Bloch-waves $|\Phi_m\rangle =|\vec{k}\rangle$.
The Laplace transform (\ref{canonic-occupation-probabilities}) of the transition matrix
is then obtained as
\begin{equation}
 \label{greensfunction}
  {\mathbf P}(s,\vec{p}-{q}) =
 \frac{\left(1-{\tilde \chi}(s)\right)}{s} \frac{1}{(2\pi)^d} 
 \int_{-\pi}^{\pi}{\rm d}k_1\ldots \int_{-\pi}^{\pi}{\rm d}k_d 
 \frac{e^{-i\vec{k}\cdot(\vec{p}-\vec{q})}}{(1- \lambda(\vec{k}){\tilde \chi}(s))} .
\end{equation}
The eigenvalues (\ref{eigvals})
for $k \rightarrow 0$ (where $k=|\vec{k}|$) then take the form
\begin{equation}
 \label{transitionlaceigs}
 \lambda(\vec{k}) \approx 1-\frac{1}{2d}k^2 .
\end{equation}
The Laplace transformed eigenvalues of the transition matrix then are obtained 
from canonic representation (\ref{greensfunction}). 
For $k$ small with (\ref{transitionlaceigs}) the Montroll-Weiss equation takes the form

\begin{equation}
 \label{PKS}
{\tilde P}(k,s) = \frac{\left(1-{\tilde \chi}(s)\right)}{s}\frac{1}{(1-{\tilde \chi}(s)\lambda(k))} = 
\frac{s^{-1}}{\left(1+\frac{{\tilde \chi}(s)}{(1-{\tilde \chi}(s))}\frac{\mu(\vec{k})}{2d}\right)}\approx 
 \frac{s^{-1}}{\left(1+\frac{{\tilde \chi}(s)}{(1-{\tilde \chi}(s))}\frac{k^2}{2d}\right)} .
 \end{equation}
We employ now the Montroll-Weiss equation (\ref{PKS}) as point 
of departure to derive a diffusional equation governing a
CTRW subordinated to a GFPP with the waiting time density described by Laplace transform 
(\ref{jump-gen-fractional-laplacetr}).
After some elementary manipulations we can rewrite the 
Montroll-Weiss Eq. (\ref{PKS}) in the Fourier-Laplace domain in the form

%
%

\begin{equation}
 \label{diff-relation}
 \begin{array}{l}
 \ds -\frac{\xi^{\alpha}}{2d} \mu(\vec{k}) {\tilde P}_{\beta,\alpha}(k,s) = (s^{\beta}+\xi)^{\alpha} \left({\tilde P}_{\beta,\alpha}(k,s)-{\tilde \Phi}^{(0)}_{\beta,\alpha}(s)\right) 
-\xi^{\alpha}{\tilde P}_{\beta,\alpha}(k,s) \\ \\  \ds
 - \frac{h^2}{2d} {\bar k}^2   {\tilde P}_{\beta,\alpha}(h{\bar k},s) \approx \left((1+\frac{1}{\xi}s^{\beta})^{\alpha}-1\right){\tilde P}_{\beta,\alpha}(h{\bar k},s) 
 +\frac{1}{s}\left(1-(1+\frac{1}{\xi}s^{\beta})^{\alpha}\right) , \hspace{1cm} h\rightarrow 0
\end{array}
\end{equation}
The first line is an exact equation which leads to the second line holding asymptotically for small $k={\bar k}h$. We introduced in the second line
a new wave vector ${\vec {\bar k}} =({\bar k}_1,..,{\bar k}_d)$. Then for any finite ${\bar k}$ we can choose $h$ sufficiently small that 
$\mu(\vec{k}) \approx k^2= ({\bar k}h)^2 \rightarrow 0$ thus
the left-hand side of the second line tends to zero when $h\rightarrow 0$. To maintain the second line of relation (\ref{diff-relation}) 
`small' requires that $\xi(h) \rightarrow \infty$
for $h\rightarrow 0$. Hence we can expand $(1+\frac{1}{\xi}s^{\beta})^{\alpha} \approx 1+ \frac{\alpha}{\xi}s^{\beta}$ to arrive at
$$
 - \frac{h^2\xi }{2d\alpha} {\bar k}^2   {\tilde P}_{\beta,\alpha}(h{\bar k},s) \approx s^{\beta} {\tilde P}_{\beta,\alpha}(h{\bar k},s) - s^{\beta-1},
 \hspace{2cm} {\rm (86.B)}.
$$
Assuming that $h^2\xi$ is kept constant when $h\rightarrow 0$ leads to the scaling 
$\xi(h) \sim h^{-2} \rightarrow \infty$. Eq. (86.B) describes the 
diffusion limit and reflects the asymptotic GFPP behavior for large dimensionless times $\xi^{\frac{1}{\beta}}(h) t \rightarrow \infty$ discussed previously 
(See Eq. (\ref{smallsgenfrac})).
The constant $\frac{h^2\xi }{2d\alpha}$ has physical dimension
$cm^2sec^{-\beta}$ and can be interpreted as a generalized fractional diffusion constant (recovering for $\beta=1$ the units $cm^2sec^{-1}$ of normal diffusion). 
We observe that the index $\alpha$ enters in Eq. {\rm (86.B)} only as a scaling parameter.

Transforming the first line of Eq. (\ref{diff-relation}) into the causal space-time 
domain 
yields (See also general relation (\ref{occupationproblaplace}))
\begin{equation}
 \label{genfracdiffusionequation}
 - \frac{\xi^{\alpha}}{2d} {\mathbf L} \cdot {\mathbf P}_{(\beta,\alpha)}(t) = 
 _0\!\mathcal{D}_t^{\beta,\alpha} \cdot {\mathbf P}_{\beta,\alpha}(t) - \xi^{\alpha} {\mathbf P}_{\beta,\alpha}(t) + 
{\mathbf I} \left\{ \xi^{\alpha}\Theta(t) - K^{(0)}_{\beta,\alpha}(t)\right\} .
\end{equation}
This relation is written in matrix form where ${\mathbf L}$ denotes the Laplacian and ${\mathbf I}$ 
the unit matrix in $\mathbb{Z}^d$ indicating the initial 
condition that the walker at $t=0$ is sitting in the origin $\vec{q}=0$. 
Eq. (\ref{genfracdiffusionequation}) describes generalized fractional diffusion in $\mathbb{Z}^d$. Its diffusion limit ($h\rightarrow 0$ and $\xi(h) \sim h^{-2}$)
is given by the space-time representation of relation {\rm (86.B)} and yields\footnote{$\Delta= \sum_{j=1}^d \frac{\partial^2}{\partial x_j^2}$ denotes the Laplace operator of the $\mathbb{R}^d$, 
$\vec{x} =h\vec{p} \in \mathbb{R}^d$ indicate rescaled quasi-continuous spatial coordinates, and $P_{(\beta,\alpha)}(\vec{q},t) = h^{d} {\cal P}_{(\beta,\alpha)}(\vec{x},t)$ with the
transition probability kernel ${\cal P}_{(\beta,\alpha)}(\vec{x},t) =
\lim_{h\rightarrow 0} \frac{1}{(2\pi)^d}\int_{-\pi h^{-\frac{1}{2}}}^{\pi h^{-\frac{1}{2}}} 
{\rm d}{\bar k}_1\dots \int_{-\pi h^{-\frac{1}{2}}}^{\pi h^{-\frac{1}{2}}} {\rm d}{\bar k}_d {\tilde P}(h{\bar k},s)e^{-i\vec{\bar k}\cdot\vec{x}}$ 
having physical units $cm^{-d}$. For more details about diffusive or long-wave limits,
see e.g. \cite{TMM-APR-ISTE2019,Gorenflo2010}.} 
$$
  \frac{h^2 \xi}{2d\alpha}\Delta {\cal P}_{(\beta,\alpha)}(\vec{x},t)  = _0 \!D_t^{\beta} \cdot {\cal P}_{(\beta,\alpha)}(\vec{x},t) - 
  \delta^{(d)}(\vec{x}) \,\frac{t^{-\beta}}{\Gamma(1-\beta)} ,\hspace{1cm} {\rm (87.B)}     
$$
where $\delta^{(d)}(\vec{x})$ denotes Dirac's $\delta$-function in $\mathbb{R}^d$ and $\vec{x} =h\vec{q} \in \mathbb{R}^d$ are the 
rescaled quasi-continuous coordinates. $ _0 \!D_t^{\beta}\cdot  P_{\beta,1}(q,t)$ indicates the Riemann-Liouville fractional derivative of order $\beta$ 
(See Appendix \ref{AppendLaplacetrafo}).
We refer the exact Eq. (\ref{genfracdiffusionequation}) to as {\it generalized fractional diffusion equation} 
having Eq. {\rm (87.B)} as `well-scaled' diffusion limit. 
Rewriting Eq. (\ref{genfracdiffusionequation}) in the Fourier-time domain yields
\begin{equation}
 \label{KGV-consistence}
 _0\!\mathcal{D}_t^{\beta,\alpha} \cdot {\hat P}_{\beta,\alpha}(k,t)    =  \xi^{\alpha}\left(1-\frac{\mu(\vec{k})}{2d}\right) {\hat P}_{(\beta,\alpha)}(k,t) +  \delta_{\vec{q},\vec{0}} \left(K^{(0)}_{\beta,\alpha}(t) - \xi^{\alpha}\Theta(t) \right)
\end{equation}
showing equivalence with the generalized fractional 
Kolmogorov-Feller equation (\ref{generalizedfracKolmogorivFeller})
by accounting for the eigenvalues $\lambda(\vec{k})= 1-\frac{\mu(\vec{k})}{2d}$ of ${\mathbf W}$ 
where we used initial condition $(\delta_{\vec{q},\vec{0}}) = {\mathbf P}(t=0)$ that the walker sits on the
origin at $t=0$. The kernels $D^{\beta,\alpha}(t)={\cal L}^{-1}\{(s^{\beta}+\xi)^{\alpha}\}$ and 
$K^{(0)}_{\beta,\alpha}(t) = {\cal L}^{-1}\{\frac{(s^{\beta}+\xi)^{\alpha}}{s}\}$ were given explicitly in Eqs. 
(\ref{en-res-mittag-leffl-gn-kernel}) and (\ref{Kzeroexplicitform}), respectively.
\\[2mm]
Now let us consider the fractional Poisson limit
$\alpha=1$ with $0<\beta<1$. Then Eq. {\rm (87.B)} takes the form 
\begin{equation}
 \label{coinciding-Metzler-Klafter}
 \frac{h^2\xi}{2d}\Delta {\cal P}_{(\beta,1)}(\vec{x},t) = _0 \!D_t^{\beta} \cdot  {\cal P}_{(\beta,1)}(\vec{x},t) -  \delta^{(d)}(\vec{x}) \,\frac{t^{-\beta}}{\Gamma(1-\beta)}  \hspace{1cm} 0<\beta<1 
\end{equation}
 Eqs. (\ref{coinciding-Metzler-Klafter}) and {\rm (87.B)} coincide with 
the fractional diffusion equation given by Metzler and 
Klafter \cite{MetzlerKlafter2004} (Eq. (10) in that paper).
%
%
\\
The fractional diffusion equation of the diffusion limit {\rm (87.B)} represents also the 
asymptotic limit $t\xi^{\frac{1}{\beta}}\rightarrow \infty$ of the 
generalized fractional diffusion equation (\ref{genfracdiffusionequation}) with the memory term 
$-\delta^{(d)}(\vec{x})\frac{t^{-\beta}}{\Gamma(1-\beta)}$ (where $\xi \rightarrow \xi'=\frac{\xi}{\alpha}$)
describing the slow power-law decay of the contribution of the initial condition (See also Eq. (\ref{diff-relation})
for $|s|$ small).
\\[2mm]
Finally we consider the standard Poisson limit $\alpha=1$ and $\beta=1$. Then Eq. 
(\ref{genfracdiffusionequation}) yields\footnote{The Fourier-Laplace domain representation
of Eq. (\ref{Poisson-recover}) is with (\ref{diff-relation}) given by: 
$ -\frac{h^2\xi}{2d}{\bar k}^2{\tilde P}_{1,1}({\bar k}h,s) = s {\tilde P}_{1,1}(h{\bar k},s) - 1$.}

\begin{equation}
 \label{Poisson-recover}
 \frac{h^2\xi}{2d}\Delta {\cal P}_{(1,1)}(\vec{x},t) = \frac{\partial }{\partial t}\left\{\Theta(t) {\cal P}_{(1,1)}(\vec{x},t) \right\} -  
 \delta^{d}(\vec{x})\delta(t)
\end{equation}
where $\frac{h^2\xi}{2d}$ indicates the diffusion constant of standard (normal-) diffusion. 
It is important here that we take into account the $\Theta(t)$-function `under the derivative'. Then by using
\begin{equation*}
\frac{\partial }{\partial t}\left(\Theta(t) {\cal P}_{(1,1)}(\vec{x},t)  \right)= \delta(t) {\cal P}_{(1,1)}(\vec{x},t) +
\Theta(t)\frac{\partial }{\partial t} {\cal P}_{(1,1)}(\vec{x},t)  = 
\delta(t) \delta^{(d)}(\vec{x}) +\Theta(t)\frac{\partial }{\partial t} {\cal P}_{(1,1)}(\vec{x},t)
\end{equation*}
with 
($P_{1,1}(q,0)=\delta_{\vec{q},\vec{0}}$), Eq. (\ref{Poisson-recover}) recovers then
Fick's second law of normal diffusion
\begin{equation}
 \label{Fick2}
 \frac{h^2\xi}{2d}\Delta {\cal P}_{(1,1)}(\vec{x},t)  = \frac{\partial }{\partial t} {\cal P}_{(1,1)}(\vec{x},t) .
\end{equation}
Finally let us consider the mean squared displacement in generalized fractional diffusion.
Then we obtain for the Laplace-Fourier transform of the mean squared displacement ${\tilde \sigma^2}(s)$ 
from the general relation
\begin{equation}
 \label{meandisplacementsGFPP}
 \begin{array}{l} \ds
 {\tilde \sigma^2}_{\beta,\alpha}(s) = -\Delta_{{\vec {\bar k}}} {\tilde P}(h{\bar k},s)\Big|_{{\bar k}=0} = h^2 {\tilde {\bar n}}_{\beta,\alpha}(s) \\ \\
 \ds \hspace{0.5cm} = 
h^2 \frac{{\tilde \chi}_{\beta,\alpha}(s)}{s (1-{\tilde \chi}_{\beta,\alpha}(s))} =
 \frac{1}{s[(\frac{s^{\beta}}{\xi}+1)^{\alpha}-1]} \approx \frac{\xi h^2}{\alpha}s^{-\beta-1} ,\hspace{0.5cm} \xi(h) \sim h^{-2} \rightarrow \infty
 \end{array} 
\end{equation}
where ${\bar n}_{\beta,\alpha}(t)$ indicates the expected number of arrivals of Eq. (\ref{Laplacetranfostepsaverge}).
For the GFPP the asymptotic behavior of the expected arrival number 
${\bar n}_{\beta,\alpha}(t)$ was also obtained in relation
(\ref{largetimesteps}). We hence obtain for the mean square displacement the behavior
\begin{equation}
 \label{tworegimesGFPP}
\ds \sigma^2_{\beta,\alpha}(t) 
\approx  \frac{h^2\xi }{\alpha \Gamma(\beta+1)} t^{\beta}  , 
\hspace{0.5cm} 0 < \beta \leq 1 \hspace{0.5cm} \alpha >0.
\end{equation}
We obtain for $0<\beta<1$ and for all $\alpha>0$ including fractional Poisson $\alpha=1$
sublinear $t^{\beta}$ power-law behavior 
corresponding to fat-tailed jump density (\ref{tlarge-fat-tailed}). 
The sublinear power law $\sigma^2(t)\sim t^{\beta}$ ($0<\beta<1$) represents the universal limit for 
$t\xi^{\frac{1}{\beta}} \rightarrow\infty$ of fat-tailed jump PDFs. Such behavior 
is well known in the literature for anomalous {\it subdiffusion} 
($0<\beta<1$) \cite{MetzlerKlafter2000}. 

In contrast for $\beta=1$ 
for all $\alpha >0$ (Erlang regime which includes the standard 
Poisson $\alpha=1$, $\beta=1$) we obtain normal diffusive behavior with linear 
increase $\sigma^2_{\beta,\alpha}(t) \approx 
\frac{h^2\xi}{\alpha} t$ of the mean squared displacement. 
This behavior can be attributed to {\it normal diffusion} \cite{MetzlerKlafter2000}.
The occurrence of universal scaling laws for large $t\xi^{\frac{1}{\beta}}\rightarrow \infty$
reflects the asymptotic universality for time 
fractional dynamics shown by Gorenflo and Mainardi \cite{GorenfloMainardi2006}.

\section{Conclusion}

We have analyzed a generalization of the Laskin fractional Poisson process, the so called 
generalized fractional Poisson process introduced for the first time to our knowledge by Polito and Cahoy 
in 2013 \cite{PolitoCahoy2013}. We developed the stochastic motions on undirected networks and 
$d$-dimensional infinite integer lattices for this renewal process.
We derived the probability 
distribution for $n$ arrivals within time interval $[0,t]$
(Eq. (\ref{Generalized-Fractional-Poisson-Distribution})) which we call the
`{\it generalized fractional Poisson distribution}' (GFPD).

The GFPP is 
non-Markovian and introduces long-time memory effects. 
The GFPP contains two index parameters $\alpha >0$ and $0< \beta \leq 1$ and a characteristic time scale 
$\xi^{-\frac{1}{\beta}}$ controlling the dynamic behavior.
The GFPP recovers for $\alpha=1$ and $0<\beta<1$ the Laskin fractional Poisson process, 
for $\beta=1$, $\alpha>0$ the Erlang process,
and for $\alpha=1$, $\beta=1$ the standard Poisson process.
We showed that for $\alpha=1$ and $0<\beta<1$ the GFPD recovers Laskin's 
fractional Poisson distribution (\ref{fractionalpoission-distribution}) and for $\alpha=1$ with $\beta=1$ 
the standard Poisson distribution. For small dimensionless observation times
 $\xi^{\frac{1}{\beta}} t \rightarrow 0$ two distinct regimes emerge, 
 $\alpha\beta >1$ with `slow' jump dynamics
and $\alpha\beta <1$ with `fast' jump dynamics with a transition at $\alpha\beta=1$ 
(See Eq. (\ref{lowest-order-in-t}), and Figures \ref{Fig_1} and \ref{Fig_2}). 

Based on the Montroll-Weiss CTRW approach we analyzed the GFPP on undirected 
networks and derived a `generalized fractional Kolmogorov-Feller equation'.
We analyzed for the resulting stochastic motion in $d$-dimensional infinite lattices 
the `well-scaled' diffusion limit of the `generalized fractional diffusion equation' 
representing a network variant of the
`generalized fractional Kolmogorov-Feller equation'.

We showed that the generalized Kolmogorov-Feller equation for $\alpha=1$ and $0<\beta<1$ 
recovers the fractional Kolmogorov-Feller equation.
For $\alpha=1$, $\beta=1$
the classical Kolmogorov-Feller equations with Fick's second law as diffusion limit
of the standard Poisson process are reproduced.
An essential feature of generalized fractional diffusion is the emergence of
subdiffusive behavior $\sigma^2(t) \sim t^{\beta}$ in the diffusion limit  governed by the same type of fractional diffusion equation ({\rm 87.B}) as occurring
with the purely fractional Poisson process ($\alpha=1$ and $0<\beta<1$). Generalized fractional diffusion turns into
normal diffusion in the Erlang regime $\beta=1$, $\alpha>0$ with $\sigma^2(t) \sim t$ 
(Eq. (\ref{tworegimesGFPP}) for $\beta=1$).
\\[2mm]
We mention that further generalizations of the fractional Poisson process are of interest. For instance we can define 
a renewal processes which 
generalize the GFPP waiting time density (\ref{jump-gen-fractional-laplacetr})
by \cite{michel-riascios-tobepublished}
\begin{equation}
 \label{furthergen}
 {\tilde \chi}_{\gamma,\beta,\alpha}(s) =  \frac{(\xi_1^{\alpha}+\xi_2)^{\gamma}}{\{(s^{\beta}+\xi_1)^{\alpha}+\xi_2\}^{\gamma}} 
 \hspace{0.5cm} 0<\beta\leq 1,\hspace{0.5cm} \alpha >0 , \hspace{0.5cm} \gamma>0 , \hspace{0.5cm} \xi_1, \xi_2 >0 , 
\end{equation}
where ${\tilde \chi}_{\gamma,\beta,\alpha}(s)|_{s=0}=1$ (normalization) and recovers for $\alpha=1$
a GFPP with $\chi_{\beta,\gamma}(t)$ waiting time density, then for $\alpha,\gamma=1$ with $0<\beta<1$ the fractional Poisson process and finally for $\alpha,\beta,\gamma=1$ 
the standard Poisson process. The advantage of generalizations such as (\ref{furthergen}) is that it introduces three index parameters 
$\alpha,\beta,\gamma$, and two time scale parameters, 
$\xi_{1,2}$ ($\xi_1$ having physical dimension $\sec^{-\beta}$, and $\xi_2$ is of dimension $\sec^{-\beta\alpha}$). 
Such generalizations 
offer
a great flexibility to fit various real-world situations such as occurring 
in various problems of anomalous diffusion, e.g. \cite{GoychukHaenggi2004} where power 
law asymptotic features occur.
The analysis of the process defined by (\ref{furthergen}) is beyond the scope of the present article and
will be presented in a follow-up paper \cite{michel-riascios-tobepublished}.
\\[2mm]
Our results suggest that the GFPP and further generalizations have huge potentials of applications 
in anomalous diffusion and transport phenomena including turbulence, non-Markovian dynamics on networks,
and in the dynamics of complex systems.

\section{Acknowledgment}
We thank Federico Polito for his valuable comments and to have drawn our attention to
Ref. \cite{PolitoCahoy2013}.

\begin{appendix}

\section{Appendix: Laplace transforms of causal distributions and fractional operators}
\label{AppendLaplacetrafo}

Here we discuss some properties of Laplace transforms that we use in the present paper. 
The Laplace transform of the $m$th time derivative of convolution
of two causal functions $h(t)\Theta(t)$ and $k(t)\Theta(t)$ is given by
\begin{equation}
 \label{convdef-aa}
 \begin{array}{l}
\ds {\cal L}\left\{\frac{d^ m}{dt^ m}\int_0^t h(v)k(t-v){\rm d}v \right\}\\ \\ \ds \hspace{0.5cm} = 
\int_{-\infty}^{\infty} \int_{-\infty}^{\infty}{\rm d}u {\rm d}v 
\Theta(u) \Theta(v) h(u) k(v)\int_{0-}^ {\infty}e^{-st} \frac{d^ m}{dt^ m}\delta(t-u-v) {\rm d}s \\ \\
\ds \hspace{0.5cm} = s^m \int_{-\infty}^{\infty} \int_{-\infty}^{\infty}{\rm d}u {\rm d}v 
\Theta(u) \Theta(v) h(u) k(v)e^{-su}e^ {-sv} 
= s^m{\tilde k}(s){\tilde h}(s) ,\hspace{1cm} m=0,1,2,\ldots  \in \mathbb{N}_0 .
\end{array}
\end{equation}
The last relation is crucial to define in our analysis
`good' fractional derivatives and integrals. 
\\[2mm]
Let us consider here property (\ref{it-follows-that}) in more details, namely
\begin{equation}
 \label{heavsdederiv}
 \begin{array}{l}
 \ds {\cal L}\left\{\frac{d^m}{dt^m}\left(\Theta(t)f(t)\right)\right\} = s^m{\tilde f}(s) \\ \\ \ds  = 
 {\cal L}\left\{\Theta(t)\frac{d^m}{dt^m}f(t)+\sum_{m=1}^m\frac{m!}{(m-k)!k!}
 \int_{0_{-}}^{\infty}e^{-st}\delta^{k-1}(t)
 \frac{d^{m-k}}{dt^{m-k}}f(t){\rm d}t \right\}
 \end{array}
\end{equation}
where is denoted $\delta^{s}(t)=\frac{d^s}{dt^s}\delta(t)$. Then integrating by parts 
these terms and using $\frac{d^{n}}{dt^n}(e^{-st}y(t))= e^{-st} (\frac{d}{dt}-s)^ny(t)$ 
we arrive at
\begin{align}
\nonumber
\int_{0_{-}}^{\infty}e^{-st}\delta^{k-1}(t)
 \frac{d^{m-k}}{dt^{m-k}}f(t){\rm d}t &= \int_{0_{-}}^{\infty} \delta(t)
 e^{-st} (-1)^{k-1}\left(\frac{d}{dt}-s\right)^{k-1} \frac{d^{m-k}}{dt^{m-k}}f(t){\rm d}t \\ \label{by-parts}
&= \left(s-\frac{d}{dt}\right)^{k-1}
 \frac{d^{m-k}}{dt^{m-k}} f(t)\Big|_{t=0} .
\end{align} 
We emphasize that the $\delta(t)$-function is entirely captured thus in (\ref{by-parts})
occur only initial values at $t=0$ (not at $t=0_{-}$).
Then by substituting (\ref{by-parts}) into (\ref{heavsdederiv}) yields the well known relation
\begin{equation}
 \label{well-known}
 {\cal L}\left\{\Theta(t)\frac{d^m}{dt^m}f(t)\right\} = s^m{\tilde f}(s)-\left(s-\frac{d}{dt}\right)^{-1}
 \left(s^m-\frac{d^m}{dt^m}\right)f(t)_{t=0} .
\end{equation}
\\[2mm]
We often will meet in our analysis the inverse Laplace transform ${\cal L}^{-1}\{s^{\gamma}\}$ ,($\gamma >0$) 
which we evaluate here in details.
To this end it is useful to introduce the {\it ceiling function} $\ceil{\gamma}$ indicating the smallest integer greater or equal to $\gamma$, i.e. for $\gamma \notin \mathbb{N}$
we have $-1<\gamma-\ceil{\gamma}< 0$ and $\gamma=\ceil{\gamma}$ if $\gamma \in \mathbb{N}$.
We can represent ${\cal L}^{-1}\{s^{\gamma}\}$
as Fourier integral, namely
\begin{equation}
 \label{inversion}
  {\cal L}^{-1}\{s^{\gamma}\} = \frac{1}{2\pi i}\int_{\sigma-i\infty}^{\sigma+i\infty}e^{st}s^{\gamma}{\rm d}s = \frac{e^{\sigma t}}{2\pi}\int_{-\infty}^{\infty} e^{i\omega t} (\sigma +i\omega)^{\gamma} {\rm d}\omega ,
 \hspace{0.5cm} s=\sigma+i\omega ,\hspace{1cm} \sigma =\Re(s) >0 .
\end{equation}
We see that for integer $\gamma \in \mathbb{N}_0$ integral (\ref{inversion}) yields $\frac{d^{\gamma}}{dt^{\gamma}}\delta(t)$ i.e. the convolution operator of the integer order derivative $\gamma$.
Let us now consider the case $\gamma >0$ and $\gamma \notin \mathbb{N}$.
Then we take into account that
\begin{equation}
\label{axiliary-rel}
 s^{\gamma} = s^{\ceil{\gamma}} s^{\gamma -\ceil{\gamma}} =  \left\{ \begin{array}{l}  \ds                                            
\frac{s^{\ceil{\gamma}}}{{\Gamma(\ceil{\gamma}-\gamma)}} \int_{-\infty}^{\infty}\Theta(t) e^{-s\tau} \tau^{\ceil{\gamma}-\gamma-1} {\rm d}\tau ,\hspace{1cm} \gamma \notin \mathbb{N} \\ \\ \ds
s^{\gamma}  \int_{-\infty}^{\infty}  e^{-s\tau} \delta(\tau){\rm d}\tau ,\hspace{1cm} \gamma \in \mathbb{N}
\end{array}\right.
\end{equation}
where it is important to notice that we write $s^{\gamma}$ in the form (\ref{axiliary-rel}) since the $\tau$-integral converges as $0<\ceil{\gamma}-\gamma<1$ and the Heaviside $\Theta(t)$-function guarantees that
the integral starts from $\tau=0$.
Then we further use the simple relation
\begin{equation}
 \label{simple-rel}
 s^{\ceil{\gamma}} e^{-s\tau} = (-1)^{\ceil{\gamma}} \frac{d^{\ceil{\gamma}}}{d\tau^{\ceil{\gamma}}} e^{-s\tau} .
\end{equation}
Then by substituting this relation into integral (\ref{axiliary-rel}) yields
\begin{equation}
\label{axiliary-rel2}
\begin{array}{l}
 \ds s^{\gamma} = \left\{ \begin{array}{l}  \ds
 \frac{ (-1)^{\ceil{\gamma}} }{{\Gamma(\ceil{\gamma}-\gamma)}} \int_{-\infty}^{\infty}\Theta(t) \tau^{\ceil{\gamma}-\gamma-1} 
 \frac{d^{\ceil{\gamma}}}{d\tau^{\ceil{\gamma}}} e^{-s\tau} {\rm d}\tau ,\hspace{1cm} \gamma \notin \mathbb{N}
 \\ \\ \ds
 (-1)^{\gamma} \int_{-\infty}^{\infty}\delta(\tau)  \frac{d^{\gamma}}{d \tau^{\gamma}} e^{-s\tau} {\rm d}\tau ,\hspace{1cm} \gamma \in \mathbb{N}
 \end{array}\right. 
 \\ \\
 \ds  (\sigma+i\omega)^{\gamma} =   \left\{ \begin{array}{l}  \ds
 \int_{-\infty}^{\infty}  e^{-(\sigma+i\omega)\tau}  \frac{d^{\ceil{\gamma}}}{d\tau^{\ceil{\gamma}}} \left\{ \Theta(\tau) 
 \frac{\tau^{\ceil{\gamma}-\gamma-1}}{{\Gamma(\ceil{\gamma}-\gamma)}}  \right\} {\rm d}\tau ,\hspace{1cm} \gamma \notin \mathbb{N} \\ \\ \ds
 \int_{-\infty}^{\infty}  e^{-(\sigma+i\omega)\tau} \frac{d^{\gamma}}{d\tau^{\gamma}}\delta(\tau){\rm d}\tau ,\hspace{1cm} \gamma \in \mathbb{N}
 \end{array}\right.
 \end{array}
\end{equation}
where in the second line $\ceil{\gamma}$ partial integrations have been performed and the boundary terms $\frac{d^{k}}{d\tau^k}(\Theta(\tau) \tau^{\ceil{\gamma}-\gamma-1}\ldots)\big|_{-\infty}^{\infty}$ are all vanishing.
In this relation is important that the Heaviside $\Theta(\tau)$ is {\it included} within the differentiation. 
With representations in (\ref{axiliary-rel2}) the Laplace inversion 
(\ref{inversion}) then yields
\begin{equation}
 \label{Laplaceinversionfracderiv}
 \begin{array}{l} 
 \ds {\cal L}^{-1}\{s^{\gamma}\} = \left\{ \begin{array}{l} \ds
 e^{\sigma t}
 \int_{-\infty}^{\infty} \underbrace{\int_{-\infty}^{\infty}\frac{e^{i\omega(t-\tau)}}{2\pi}{\rm d}\omega}_{=\delta(t-\tau)} 
 e^{- \sigma\tau} \frac{d^{\ceil{\gamma}}}{d\tau^{\ceil{\gamma}}} \left\{ \Theta(\tau) \frac{\tau^{\ceil{\gamma}-\gamma-1}}{{\Gamma(\ceil{\gamma}-\gamma)}}  \right\} {\rm d}\tau 
 ,\hspace{1cm} \gamma \notin \mathbb{N} \\ \\ \ds
 e^{\sigma t}
 \int_{-\infty}^{\infty} \underbrace{\int_{-\infty}^{\infty}\frac{e^{i\omega(t-\tau)}}{2\pi}{\rm d}\omega}_{=\delta(t-\tau)} 
 e^{- \sigma\tau} \frac{d^{\gamma}}{d\tau^{\gamma}}\delta(\tau) {\rm d}\tau ,\hspace{1cm} \gamma \in \mathbb{N}
 \end{array}\right.
 \\ \\ \ds 
 {\cal L}^{-1}\{s^{\gamma}\} = \left\{\begin{array}{l} \frac{d^{\ceil{\gamma}}}{dt^{\ceil{\gamma}}} \ds
 \left\{ \Theta(t) \frac{t^{\ceil{\gamma}-\gamma-1}}{{\Gamma(\ceil{\gamma}-\gamma)}}  \right\}  , \hspace{1cm} \gamma \notin \mathbb{N} \\ \\ \ds
  \frac{d^{\gamma}}{dt^{\gamma}}\delta(t)  , \hspace{1cm} \gamma \in \mathbb{N}. \end{array} \right.
 \end{array}
\end{equation}
For fractional $\gamma \notin \mathbb{N}$ we identify the kernel of 
the {\it Riemann-Liouville fractional derivative} of order 
$\gamma$, e.g. \cite{OldhamSpanier1974,MillerRoss1993} and many others, 
and for $\gamma \in \mathbb{N}$ we get the distributional kernel producing integer derivatives of order $\gamma$. 
The Riemann-Liouville fractional derivative acts on causal functions $f(t)=\Theta(t)f(t)$ as 
\begin{equation}
 \label{fractderiv}
 \begin{array}{l}
 \ds {\cal L}^{-1}\{s^{\gamma}\} \cdot f(t)\Theta(t) =: _0\!D_t^{\gamma} f(t) = \frac{d^{\ceil{\gamma}}}{dt^{\ceil{\gamma}}}\int_{-\infty}^{\infty}
 \left\{ \Theta(t-\tau) 
 \frac{(t-\tau)^{\ceil{\gamma}-\gamma-1}}{{\Gamma(\ceil{\gamma}-\gamma)}}  \right\} f(\tau)\Theta(t) {\rm d}\tau \\ \\
\ds _0\!D_t^{\gamma} f(t) = \frac{1}{{\Gamma(\ceil{\gamma}-\gamma)}}  \frac{d^{\ceil{\gamma}}}{d\tau^{\ceil{\gamma}}} \int_0^t (t-\tau)^{\ceil{\gamma}-\gamma-1} f(\tau){\rm d}\tau .
 \end{array}
\end{equation}
We emphasize that (\ref{Laplaceinversionfracderiv}) requires causality and 
definition of the Laplace transform (\ref{LaplaceDef}) that captures the entire non-zero contributions 
of the causal distribution
with lower integration limit $0_{-} =\lim_{\epsilon\rightarrow 0+}-\epsilon$ where 
all boundary terms at $0_{-}$ are vanishing.
The same result (\ref{Laplaceinversionfracderiv}) for the Riemann-Liouville fractional 
derivative kernel is obtained in the following short way
\begin{equation}
 \label{part1}
 \begin{array}{l}
\ds {\cal L}^{-1}\{s^{\gamma}\}=  {\cal L}^{-1} \{s^{\ceil{\gamma}} s^{\gamma -\ceil{\gamma}}\} = 
  e^{\sigma t}\left(\sigma+\frac{d}{dt}\right)^{\ceil{\gamma}}\left(\sigma+\frac{d}{dt}\right)^{\gamma-\ceil{\gamma}} \delta(t) 
  ,\hspace{1cm} \gamma >0 , \gamma \notin \mathbb{N}
  \\ \\ \ds
 \hspace{0.5cm}  =
  e^{\sigma t}\left(\sigma+\frac{d}{dt}\right)^{\ceil{\gamma}}
 \int_{-\infty}^{\infty}\frac{{\rm d}\omega}{(2\pi)}e^{i\omega t}
 (\sigma+i\omega)^{\gamma-\ceil{\gamma}}  \\ \\ \ds \hspace{0.5cm} =
 e^{\sigma t}\left(\sigma+\frac{d}{dt}\right)^{\ceil{\gamma}} \left\{ e^{-\sigma t}\Theta(t)  
 \frac{t^{\ceil{\gamma}-\gamma-1}}{(\ceil{\gamma}-\gamma -1)!}\right\} = 
 \frac{d^{\ceil{\gamma}}}{dt^{\ceil{\gamma}}} \left(\Theta(t) 
 \frac{t^{\ceil{\gamma}-\gamma-1}}{\Gamma(\ceil{\gamma}-\gamma)} \right) .
 \end{array}
\end{equation} 
In the same way as above one can derive the Laplace inversion of $s^{-\gamma}$ , $\gamma>0$ 
which yields
\begin{equation}
 \label{Lplaceinversionnegativegamma}
 {\cal L}^{-1}\{s^{-\gamma}\}= \Theta(t) \frac{t^{\gamma-1}}{\Gamma(\gamma)} ,\hspace{1cm}  \gamma >0
\end{equation}
as a fractional generalization of integration and indeed can be identified with the Riemann-Liouville fractional 
integral kernel of order $\gamma$, e.g. \cite{OldhamSpanier1974,MillerRoss1993}.


%
%
\end{appendix}


\end{document}